\documentclass[12pt]{article}
\usepackage{xspace}

\usepackage{amssymb}
\usepackage{epsf}
\usepackage{psfrag}
\newcommand{\NPB}[3]{\emph{ Nucl.~Phys.} \textbf{B#1} (#2) #3}
\newcommand{\PLB}[3]{\emph{ Phys.~Lett.} \textbf{B#1} (#2) #3}
\newcommand{\PRD}[3]{\emph{ Phys.~Rev.} \textbf{D#1} (#2) #3}

\newcommand{\HPA}[3]{\emph{ Helv.~Phys.~Acta} \textbf{#1} (#2) #3}
\newcommand{\AP}[3]{\emph{ Ann.~Phys.} \textbf{#1} (#2) #3}

\newcommand{\JETP}[3]{\emph{ JETP Lett.} \textbf{#1} (#2) #3}

\newcommand{\nc}{\newcommand}

\def\simlt{\stackrel{<}{{}_\sim}}
\def\simgt{\stackrel{>}{{}_\sim}}

\nc{\beq}{\begin{equation}}
\nc{\eeq}{\end{equation}}
\nc{\beqa}{\begin{eqnarray}}
\nc{\eeqa}{\end{eqnarray}}
\nc{\bea}{\begin{eqnarray}}
\nc{\eea}{\end{eqnarray}}
\nc{\ra}{\rightarrow}
\nc{\lsim}{\begin{array}{c}\,\sim\vspace{-21pt}\\< \end{array}}
\nc{\gsim}{\begin{array}{c}\sim\vspace{-21pt}\\> \end{array}}

\nc{\LL}{L}
\nc{\vv}{\tilde{v}}

\makeatletter
\@addtoreset{equation}{section}
\makeatother

\title{
\vspace*{-1.3cm}
\begin{flushright}
\normalsize{
ANL-HEP-PR-04-87 \\
EFI-04-31 \\
FERMILAB-PUB-04-264-T\\
UAB-FT-572
}
\end{flushright}
\vspace{1.5cm}
\Large
\textbf{\sc Electroweak Baryogenesis and New TeV Fermions}
\vspace*{.5cm}
\author{\large \textbf{M.~Carena~$^a$}, \textbf{A.~Megevand~$^b$} \\ \\
\textbf{M.~Quir\'os~$^{b,\,c}$} and \textbf{C.E.M.~Wagner~$^{d,\,e}$}\\ [0.5cm]
$^a$\normalsize\emph{Fermi National Accelerator Laboratory,
P.O. Box 500, Batavia, IL 60510, USA} \\
$^b$\normalsize\emph{Theoretical Physics Group,
IFAE/UAB, E-08193 Bellaterra (Barcelona) Spain} \\
$^c$\normalsize \emph{Instituci\'o Catalana de Recerca i Estudis Avan\c{c}ats
(ICREA)}\\
$^d$\normalsize\emph{HEP Division, Argonne National Laboratory,
9700 S. Cass Ave.,
Argonne, IL 60439, USA} \\
$^e$\normalsize\emph{Enrico Fermi Institute, Univ. of Chicago, 5640
Ellis Ave., Chicago, IL 60637, USA}}}

\begin{document}
\setcounter{page}{0}
\maketitle
\vspace*{1cm}
\begin{abstract}
New fermions, strongly coupled to the Standard Model Higgs boson,
provide a well motivated extension of the Standard Model (SM).  In
this work we show that, once new physics at heavier scales is added to
stabilize the Higgs potential, such an extension of the SM can
strengthen the first order electroweak phase transition and make the
electroweak baryogenesis mechanism feasible.  We propose
a SM extension with TeV Higgsinos, Winos and Binos that satisfy the
following properties: {\bf a)} The electroweak phase transition is
strong enough to avoid sphaleron erasure in the broken phase for
values of the Higgs mass $m_H\simlt 300$~GeV; {\bf b)} It provides
large CP-violating currents that lead to the observed baryon asymmetry
of the Universe for natural values of the CP-violating phase; {\bf c)}
It also provides a natural Dark Matter candidate that can reproduce
the observed dark matter density; {\bf d)} It is consistent with
electroweak precision measurements; {\bf e)} It may arise from a
softly broken supersymmetric theory with an extra (asymptotically
free) gauge sector; {\bf f)} It may be tested by electron electric
dipole moment experiments in the near future.
\end{abstract}

\thispagestyle{empty}
\newpage

\setcounter{page}{1}

\baselineskip18pt

\section{\sc Introduction}
\label{introduction}

In spite of all our recent progress in the understanding of physics at
the electroweak scale, the source of Dark Matter (DM) and the origin
of the matter-antimatter asymmetry~\cite{sakharov} (BAU) still remain
unclear. It is today well understood that the solution to either of
these problems requires physics beyond the Standard Model (SM).

On the one hand if the Standard Model baryon number violating
interactions~\cite{anomaly,sphalerons,sphalT} cease to be in
equilibrium in the bubbles of the broken phase, the matter-antimatter
asymmetry may be generated at the electroweak phase transition via the
mechanism of electroweak baryogenesis~\cite{EWBGreviews,ckn}. For such
a mechanism to be realized in nature a strongly first order
electroweak phase transition is required.  However the phase
transition in the Standard Model for values of the Higgs mass
consistent with the LEP bounds is a crossover~\cite{SMpt} and hence
any baryon asymmetry generated at the weak scale would be
erased. Moreover, the sources of CP-violation are insufficient to
generate a baryon number consistent with the one observed in
nature~\cite{Gavela}.

Electroweak Baryogenesis remains nevertheless as an attractive
possibility in models of physics beyond the Standard Model at the weak
scale. It has been shown that the minimal supersymmetric standard
model (MSSM) is consistent with this mechanism provided one of the
superpartners of the top quark is lighter than the top quark and the
Higgs boson is lighter than $\sim$120 GeV~\cite{CQW}--\cite{LR}.  This
mechanism also demands the presence of charginos and neutralinos at
the weak scale, which provide the necessary CP-violating
sources~\cite{HN}--\cite{CQSW} and also a natural Dark Matter
candidate.  Electroweak Baryogenesis may also be realized in the
next-to minimal supersymmetric extension of the Standard Model (NMSSM)
where some of the MSSM constraints can be relaxed. In particular there
are modifications to the tree-level effective
potential~\cite{Geraldine} that may ensure a strongly first order
phase transition~\cite{Pietroni:in}--\cite{Menon:2004wv} without a
light stop.

On the other hand the Standard Model does not provide any natural
source for the observed Dark Matter. Neutrinos are too light to give
any sizeable contribution and there is no evidence of the possible
existence of sufficient Jupiter-like, baryonic objects. Moreover,
recent WMAP measurements~\cite{Spergel:2003cb}-\cite{Tegmark:2003ud}
exclude the presence of a significant baryonic contribution to the
observed Dark Matter density. The natural candidates for the source of
Dark Matter are new, neutral, stable, weakly interacting particles
with masses of the order of the weak scale. These particles lead
naturally to a relic density of the order of the critical density and
appear in many models beyond the Standard Model. In particular they
are present in models of softly broken supersymmetry at the TeV
scale. The lightest supersymmetric particle in these models tends to
be neutral and its stability is ensured by a Parity symmetry, $R_P$,
which also ensures the proton stability.  Acceptable values of the
Dark Matter density and a successful realization of the mechanism of
electroweak baryogenesis may be simultaneously obtained in minimal
supersymmetric models, in certain phenomenologically interesting
regions of the parameter
space~\cite{Balazs:2004bu},~\cite{Menon:2004wv}.

In the previous models, as well as in all successfully considered
scenarios, the strengthening of the phase transition proceeds from the
existence of new, extra scalars in the theory, while the CP-violating
and Dark Matter sources proceed from new fermion fields. The common
lore from all previous works was that the presence of extra bosons was
a necessary requirement to induce a strong enough first order
electroweak phase transition. In fact it is currently understood that
bosons coupled to the Higgs field $\phi$ with coupling $h$ favor a
first order phase transition: they create a cubic term in the Higgs
effective potential $\sim (h\phi)^3$ either at the tree level, as in
the NMSSM, or by its contribution to the one-loop thermal effective
potential $\sim (h\phi)^3\, T$. On the other hand fermions do not give
rise to any cubic term in the high temperature expansion in powers of
$h\phi/T$ of the thermal integrals and hence they were neither
believed to give rise to a barrier between the symmetric $\phi=0$ and
broken $\phi\neq 0$ phases nor to trigger a first order phase
transition. In this paper we will prove that while the latter
statement remains true for weakly coupled fermions $h\ll 1$ it is not
for strongly coupled (but still perturbative) ones $h\simgt 1$ that
can indeed induce a strongly first order phase transition consistent
with electroweak baryogenesis.

In this work we shall first show that in Standard Model extensions
with extra fermions strongly coupled to the Higgs field the first
order phase transition may be sufficiently strengthened in order to
avoid erasure of the baryon asymmetry in the broken phase. We shall
analyze in detail a simple model, which can be considered as a
particular realization of split supersymmetry~\cite{Giudice}, where
the standard supersymmetric relations between the Yukawa and gauge
couplings are not fulfilled. We shall stress, however, that in such a
model, the physical vacuum becomes unstable and therefore the strength
of the electroweak phase transition may not be properly defined
without an ultraviolet (UV) completion of this model, that includes
the presence of heavier, stabilizing fields. An example of such fields
may be provided by softly broken supersymmetry, although other
extensions are possible.

We shall show that this low-energy effective theory, with Higgsinos
and gauginos strongly coupled to the Higgs, may arise from a soft
supersymmetry breaking model, based on a gauge extension of the
Standard Model gauge group, with new (asymptotically free) gauge
interactions that become strong at the TeV scale, and are responsible
for the strong Yukawa couplings of Higgsinos and gauginos to the SM
Higgs field. This gauge extension of the MSSM provides a UV completion
of the model analyzed in this paper and allows for large Higgs masses.

The article is organized as follows. In section~\ref{phase} we present
the general ideas leading to the strengthening of the phase transition
in the presence of strongly coupled fermions, and the need for the
presence of stabilizing fields. In section~\ref{higgsinos} the phase
transition for the Standard Model extension containing Higgsinos,
Winos and Binos strongly coupled to the Higgs field is worked out in
detail. We show that a strong enough first order phase transition can
be accommodated even with a heavy Higgs, $m_H\simlt 300$ GeV. In
section~\ref{cp} the CP-violating currents induced by the charginos
that lead to the observed baryon asymmetry of the universe for natural
values of CP-violating phases are presented. We show that in order to
reproduce the WMAP results on the BAU the CP-violating phases must be
$\mathcal O(1)$ if all squarks are heavy enough to be decoupled from
the thermal bath. Otherwise if some light squark (e.g.~the
right-handed stop) remains in the spectrum, the CP-violating phases
can be as small as a few times $10^{-3}$. In section~\ref{edm} the
two-loop contributions to the electron electric dipole moment from the
charginos and neutralinos in our model (in the absence of light
squarks) are evaluated, assuming that all relevant CP-violating
effects are associated with the new fermions.  For the values of the
parameters satisfying all other requirements the generated electric
dipole moment is below the present experimental bound, although the
model may be tested in the future if experimental bounds improve by a
few orders of magnitude.  In section~\ref{dark} we discuss the Dark
Matter constraints in our scenario, whereas compatibility of the
strongly coupled fermions with electroweak precision measurements is
considered in section~\ref{precision}. A natural region in the space
of parameters is found where all requirements are fulfilled. A
discussion on a possible UV completion of the model is presented in
section~\ref{strongc}. Finally we reserve section~\ref{conclusions}
for our conclusions.

\section{\sc Phase Transition and TeV Fermions}
\label{phase}

The finite temperature effective potential of the Higgs field $\phi$
is, by definition, the free-energy associated with $\phi$.  The
one-loop, finite temperature contribution to the free-energy density
is given by
\begin{equation}
{\mathcal{F}}_1(\phi,T) = 
\sum_i \frac{g_i T^{4}}{2\pi ^{2}}I_{\mp }\left( \frac{m_i(\phi)}{T}\right) ,
\label{bucle}
\end{equation}
where $g_i$ is the number of degrees of freedom (d.o.f.) of the
particle, $m_i(\phi)$ is the Higgs-dependent particle mass, and
\begin{equation}
I_{\mp }\left( x\right) =\pm \int_{0}^{\infty }dy\, y^{2}\log \left(
1\mp e^{- \sqrt{y^{2}+x^{2}}}\right) ,
\end{equation}
where $I_-(x)$ [$I_+(x)$] stands for the contribution from bosons
[fermions]. In this section we will consider for simplicity masses of
the form $m^{2}(\phi) =\mu^{2} +h^{2}\phi ^{2}$, where $h$ is the
Yukawa coupling and $\mu$ is a constant mass parameter~\footnote{Mass
eigenvalues are in general more complicated functions of $\phi$ as it
will be the case in the model considered in section~\ref{higgsinos}.}.
For large masses ($m/T\gg 1$) the functions $I_{\mp}$ are
exponentially small, which means that heavy species are decoupled from
the thermal plasma.  For small masses ($m/T\ll 1$) $I_{\mp}$ can be
expanded in a power series of $m/T$.  This is the case of the minimal
standard model. In fact adding this expansion to the zero-temperature
effective potential in the SM one obtains the well known expression
for the free-energy density
\begin{equation}
{\cal{F}}_{\rm SM} \left(\phi,T\right) 
=-\frac{\pi ^{2}}{90}g_{\ast }T^{4}+V_{\rm SM}\left( \phi
,T\right) ,
\label{feffsm}
\end{equation}
where the first term comes from the entropy density contribution of
relativistic particles, with $g_{\ast }$ the number of effectively
light species in the plasma~\cite{Kolb:vq} ($ g_{\ast }\simeq 107$ for
the SM), and the second term is the field dependent effective
potential
\begin{equation}
V_{\rm SM}\left( \phi ,T\right) =D\left( T^{2}-T_{0}^{2}\right) \phi
^{2}-ET\phi ^{3}+ \frac{\lambda _{T}}{4}\phi ^{4}.  \label{veff}
\end{equation}
In the SM the parameters of Eq.~(\ref{veff}) are well known and given
by~\cite{Linde}
\begin{eqnarray}
D &=&\frac{1}{8v^{2}}\left( 2m_{W}^{2}+m_{Z}^{2}+2m_{t}^{2}\right)
,\quad E=
\frac{1}{6\pi v^{3}}\left( 2m_{W}^{3}+m_{Z}^{3}\right) , \nonumber\\
T_{0}^{2} &=&\frac{1}{4D}\left( m_{H}^{2}-8Bv^{2}\right) ,\quad
B=\frac{3}{64\pi ^{2}v^{4}}\left( 2m_{W}^{4}+m_{Z}^{4}-4m_{t}^{4}\right) ,
\nonumber\\
\lambda _{T} &=&\lambda -\frac{3}{16\pi ^{2}v^{4}}\left(
2m_{W}^{4}\ln \frac{m_{W}^{2}}{a_{B}T^{2}}+m_{Z}^{4}\ln
\frac{m_{Z}^{2}}{a_{B}T^{2}}
-4m_{t}^{4}\ln \frac{m_{t}^{2}}{a_{F}T^{2}}\right) ,\nonumber \\
\lambda &=&\frac{m_{H}^{2}}{2v^{2}},\quad \ln a_{B}=3.91,\quad \ln
a_{F}=1.14.
\end{eqnarray}
where the vacuum expectation value of the Higgs at zero temperature is
normalized to $v\simeq 246$ GeV.

A free energy of the form of the SM one, Eq.~(\ref{feffsm}), leads to
a first order phase transition at a critical temperature given by
\begin{equation}
T_{c}=\frac{T_{0}}{\sqrt{1-E^{2}/\lambda _{T_{c}}D}}.
\end{equation}
The value of the Higgs field at the minimum of the potential is
\begin{equation}
\phi _{m}\left( T\right) =\frac{3ET}{2\lambda _{T}}\left[
1+\sqrt{1-\frac{8}{9}\frac{\lambda _{T}D}{E^{2}}\left(
1-\frac{T_{0}^{2}}{T^{2}}\right) }\right] ,  \label{fim}
\end{equation}
and the order parameter at the critical temperature is given by $\phi
_{c}/T_c\equiv\phi _{m}\left( T_{c}\right)/T_c =2E/\lambda _{T_{c}}$.

Notice that the parameter $E$ is very small because only the
(transverse) gauge bosons contribute to it and, as a consequence,
$\phi _{c}$ is much smaller than $T_c$ (unless $\lambda\simlt 2E$) and
the phase transition is very weakly first-order~\footnote{For $E=0$
the transition becomes second order in the one-loop
approximation.}. Moreover, for physical values of the Higgs mass, the
small value of $\phi_c/T_c$ causes perturbation theory to break down
and only non-perturbative calculations become reliable.  To overcome
this problem SM extensions containing extra bosons strongly coupled to
the Higgs sector have been considered in the
literature~\cite{EWBGreviews}. In general these bosons would
contribute to the parameter $E$ and would strengthen the first order
phase transition. In this paper we will prove that a similar effect is
produced by fermions strongly coupled to the Higgs sector, even if in
the high temperature regime they do not contribute to the cubic term
in the effective potential.

In general we will consider particle species that contribute to the
one-loop effective potential as in Eq.~(\ref{bucle}), not light enough
for the validity of the power expansion of $I_{\mp }$ but not
necessarily so heavy as to consider them to be decoupled in the
typical range of temperatures of the electroweak phase transition.
The $\phi$-dependent part of the free energy density would then be
given by
\begin{equation}
\label{freeen}
\mathcal{F}(\phi,T) = \mathcal{F}_{\rm SM}\left(\phi ,T\right) \pm
\sum_i g_i V(m^2_i(\phi))
+T^{4}\sum_i g_{i}I_{\mp }\left[m_{i}\left(\phi \right) /T\right]
/2\pi ^{2},
\end{equation} 
where the first term is given by Eq.~(\ref{feffsm}), the last one is
the finite temperature contribution of the new, heavy particles,
$V(m^2_i)$ is the zero temperature contribution, and the plus and
minus signs in front of $V(m^2_i)$ correspond to bosons and fermions,
respectively.  The zero-temperature one-loop effective potential
$V(m^2(\phi))$ is given by
\begin{equation}
V(m^2(\phi))=\frac{1}{64\pi^2}\,m^4(\phi)\log m^2(\phi)+P(\phi)
\label{vimi}
\end{equation}
where $P(\phi)$ is a polynomial in $\phi$ that contains quadratic and quartic
terms with coefficients that depend 
on the renormalization conditions~\cite{giudice}. 
By imposing the renormalization conditions already used in the SM, 
Eq.~(\ref{veff}), in particular 
that the tree-level values of the minimum and Higgs mass
are not shifted by radiative corrections,~i.e. 
\begin{equation}
\left.\frac{d V}{d\phi}\right|_{\phi=v}= 0 , \;\;\;\;\;\;\;\;\;
\left.\frac{d^2 V}{d\phi^2}\right|_{\phi=v} = m_H^2
\end{equation}
we obtain 
\begin{equation}
P(\phi)=\frac{1}{2}\alpha\, \phi^2+\frac{1}{4}\beta\,\phi^4
\end{equation}
with
\begin{eqnarray}
\alpha&=&\frac{1}{64\pi^2}\left\{\left(-3\,\frac{\omega \omega'}{v}+
\omega^{\prime\, 2}+\omega\omega''\right)\log\omega-\frac{3}{2}
\frac{\omega\omega'}{v}+\frac{3}{2}\omega^{\prime\, 2}+\frac{1}{2}
\omega\omega''\right\}\nonumber\\
\beta&=&\frac{1}{128\pi^2 v^2}\left\{2\left(\frac{\omega\omega'}{v}-
\omega^{\prime\, 2}-\omega\omega''\right)\log\omega
+\frac{\omega\omega'}{v}-3\omega^{\prime\, 2}-\omega\omega''\right\}
\end{eqnarray}
where we are using the notation: $\omega=m^2(v)$,
$\omega'=\left.\frac{dm^2(\phi)}{d\phi}\right|_{\phi=v}$, and so
on. For the case where the $\phi$ dependence of the mass eigenvalue is
$m^2(\phi)=\mu^2+h^2\phi^2$ the potential (\ref{vimi}) has the
familiar expression
\begin{equation}
V(m^2(\phi))=\frac{1}{64\pi^2}\,\left[ m^4(\phi)
\left(\log\left(\frac{m^2(\phi)}{m^2(v)}\right)
-\frac{3}{2}\right)+2m^2(\phi)m^2(v) 
\right] .
\end{equation}

Let us first stress that, unless the Higgs is heavy, strongly coupled
fermions may create a problem of vacuum stability at scales close to
the electroweak scale. This can be easily understood from the fact
that the tree-level quartic coupling, defined as the coefficient of
the quartic term in the effective potential, is given by $m_H^2/8 v^2$
and the radiative corrections are proportional to the fourth power of
the Yukawa coupling, $h$, and to the number of degrees of freedom. As
we shall demonstrate, a relevant effect may only be obtained for a
value of the number of degrees of freedom times $h^4$ larger than
$\mathcal O(10)$.  For such values vacuum stability occurs at scales
of the order of TeV and therefore the presence of new, stabilizing
fields is necessary in order to define a consistent low-energy
effective theory.

An efficient way of stabilizing the potential in the presence of
strongly coupled fermions is to assume the presence of heavy bosonic
degrees of freedom with similar couplings and number of degrees of
freedom.  For simplicity, let us here assume that the fermions have a
dispersion relation $m_f^2(\phi) = \mu_f^2 + h^2 \phi^2$, and a number
of degrees of freedom $g$, and there are bosonic, stabilizing fields
with a dispersion relation $m_S^2(\phi) = \mu_S^2 + h^2 \phi^2$ with
equal number of degrees of freedom.  Then, taking into account only
the radiative corrections associated with these heavy fields, the
maximum value of $\mu_S$ consistent with vacuum stability may be
obtained from the condition of a positive quartic coupling at scales
much larger than $v$, and it is given by
\begin{equation}
\mu_S^2 \leq \; \exp\left( \frac{m_H^2 8 \pi^2}{g \; h^4 v^2} \right)
m_f^2(v) - h^2 v^2
\label{bosonstab}
\end{equation}
Observe that for heavy Higgs bosons and/or weakly coupled fermions
$\mu_S$ becomes much larger than the weak scale. If, however, $h$
takes large values and $m_H$ becomes light then $\mu_S$ approaches
$\mu_f$ and the effect of stabilizing fields is to cancel the zero
temperature contribution of fermions plus giving additional finite
temperature contributions.

In order to get an understanding of the effects to be expected by the
presence of these new particles, let us first consider a fermion
particle with a mass, $m(\phi) =h\,\phi$, much larger than $T_c$ for
$\phi\simeq\phi_c$. If this effect is obtained for large values of the
Yukawa coupling, then the sum of the effects of the fermions and the
stabilizing fields at zero temperature is small and, in this extreme
case that maximizes the contribution of fermions (and that of the
stabilizing fields) to the phase transition, we can ignore the zero
temperature contributions.  Then, in the symmetric phase the species
is light and contributes to $g_{\ast }$, but in the broken-symmetry
phase it is heavy and approximately decouples from the thermal
plasma. Usually such a decoupling species would transfer its entropy
to the thermal bath, causing a temperature rise.  During a phase
transition we would naively expect that the effect of such a reheating
is to {\it delay} the appearance of the true vacuum, to decrease the
critical temperature and subsequently to increase the value of
$\phi_c/T_c$.

More quantitatively, at constant $g_{\ast }$ the critical temperature
is given by the condition $V\left(\phi _{m}\left(T'_{c}\right)
,T'_{c}\right) =0$, corresponding to degenerate minima of
$V$~\footnote{We are normalizing the total effective potential as
$V(0,T)=0$}. As we will show below, this condition changes if the
number of light degrees of freedom is different in the two phases,
$\Delta g_{\ast }\equiv g_{\ast\;
\textrm{symmetric}}-g_{\ast\;\textrm{broken}} \neq 0$.  In our
example, the decoupling particle has a mass $m(\phi) = h \phi$, and
its contribution to the free-energy density in the broken phase
vanishes, while in the symmetric phase is equal to $-\frac{\pi
^{2}}{90}\Delta g_{\ast }T^{4}$, where $\Delta g_{\ast}$ is its number
of degrees of freedom.  The condition of degenerate minima of
$\mathcal{F}$, Eq.~(\ref{freeen}), then gives~\cite{gw81,qcd}
\begin{equation}
V_{\rm SM}\left( \phi _{m}\left( T_{c}\right) ,T_{c}
\right) = 
-\frac{\pi ^{2}}{90}\Delta g_{\ast }T_{c}^{4}.
\label{tcprima}
\end{equation}
This condition is attained at a lower critical temperature,
$T_{c}^{\,\prime }> T_{c}$. Moreover we have that $\phi _{c}=\phi
_{m}\left( T_{c}\right) >\phi _{c}^{\prime}$ and the phase transition
is more strongly first-order. It was already noticed in
Ref.~\cite{ariel} that $\Delta g_{\ast } \neq 0$ in the context of the
electroweak phase transition could be important for baryogenesis.

We can now estimate $\Delta \left( \phi _{c}/T_{c}\right) \equiv \phi
_{c}/T_{c}- \phi_{c}^{\prime }/T_{c}^{\,\prime }$ by noticing that the
value of the effective potential~(\ref{veff}) at the minimum
$\phi_{m}$ can be written as
\begin{equation}
\frac{V_{\rm SM}\left(\phi _{m}\left( T\right),T\right)
}{T^{4}}=\frac{\lambda }{4} \left( \frac{\phi _{m}(T)}{T}\right)
^{3}\left( \frac{\phi^{\prime}_{c}}{T^{\prime}_{c}}
-\frac{ \phi _{m}(T)}{T}\right) ,
\end{equation}
where we have made the approximation $\lambda _{T}\simeq\lambda$.
Eq.~(\ref{tcprima}) implies that
\begin{equation}
\left( \frac{\phi _{c}}{T_{c}}\right)
^{3}\Delta \left( \frac{\phi _{c}}{T_{c}}\right) =\frac{2\pi
^{2}\Delta g_{\ast }}{45\lambda }.  \label{deltafi}
\end{equation}
For the order parameter to increase from $\phi^{\prime}
_{c}/T^{\prime}_{c}=2E/\lambda\ll 1$ to the value $\phi _{c}/T_{c}
\simgt 1$, necessary to preserve the baryon
asymmetry~\cite{EWBGreviews}, we need $\Delta g_{\ast}\simgt 45\lambda
/2\pi ^{2}\simeq 0.25\, (m_H/115\, {\rm GeV})^2$. So it seems that
particles with few d.o.f. will produce an effect. Conversely given
$\Delta g_{\ast }$, the bound on the Higgs mass is relaxed from
$(m_H/v)^2 <4E$ to
\begin{equation}
(m_H/v)^2 <4E+4\pi^{2}\Delta g_{\ast }/45. 
\end{equation}
Notice that if the new value $\phi _{c}$ is $\sim T_c$, then the
perturbative (one-loop) approach is fully justified even for small
values of $E$.

It should be noticed that Eqs.~(\ref{tcprima})--(\ref{deltafi}) give
only an estimate of $T_{c}$ and $\phi_{c}$. In fact, the effective
number of light d.o.f. varies continuously from $g_{\ast }$ to
$g_{\ast }-\Delta g_{\ast }$, as $\phi$ goes from $0$ to
$\phi_{c}$. Therefore we have a function $g_{\ast }\left( \phi
\right)$ that contributes to $\mathcal{F}\left(\phi,T \right) $, and
the correct value of the vacuum expectation value (VEV) is obtained by
minimizing the complete free energy, Eq.~(\ref{freeen}).  Once the
dependence of $g_{\ast}$ on $\phi$ is taken into account, we cannot
use analytical approximations and we need to resort to numerical
calculations. Observe that in the limit studied above fermions and
bosons gave equally important contributions to the phase transition
strength and the number of degrees of freedom is the sum of the one
associated to the fermions and that of stabilizing fields. In the
rest of this work we will consider bosons that are heavier than the 
fermions, and therefore lead to a smaller finite temperature contribution 
than in the example above.

\begin{figure}[htb]
\psfrag{mu}[][tl]{$\mu$ [GeV]}\psfrag{g}[][l]{$g$} \psfrag{h}[][r]{$h$}
\centering \epsfysize=6cm \leavevmode \epsfbox{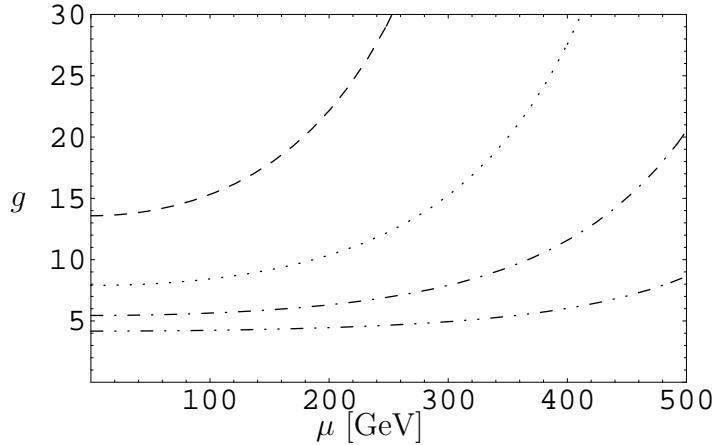}
\caption{\it Curves of constant $\phi_c/T_c=1$ and $m_H = 120$ GeV,
for a fermion with mass $m^2=\mu^2+h^2\phi^2$ and $g$ degrees of
freedom.  From top to bottom the curves correspond to $h=1.5$, $2$,
$2.5$, and $3$.}
\label{fermiones}
\end{figure}

We will now consider adding a fermion particle with mass $m^2(\phi)
=\mu^2+h^2\,\phi^2$ to the SM with a Higgs mass $m_{H}=120$~GeV, but
we will retain the effects of the heavy particles in the broken
phase. We consider only fermion species and their (heavier)
stabilizing fields since, as explained above, the effect of bosons on
the phase transition has been extensively studied in the
literature~\cite{EWBGreviews},~\cite{HN}--\cite{CQSW}.  In
Fig.~\ref{fermiones} we plot the number of degrees of freedom $g$ that
give $\phi_c/T_c=1$ as a function of the mass parameter $\mu$, for
different values of the Yukawa coupling $h$. The invariant mass of the
stabilizing fields has been set to their maximum value consistent with
vacuum stability, Eq.~(\ref{bosonstab}). As anticipated, for $\mu
\simeq 0$, and for large values of the Yukawa couplings, only a small
number of degrees of freedom are necessary in order to obtain a
strongly first order phase transition.

A minimal Standard Model extension (i.e.~the introduction of a single
species) is possible with bosons but not with fermions, since the SM
Higgs is an $SU(2)$ doublet. In order to construct an invariant Yukawa
Lagrangian the simplest possibilities are, either doublet and singlet
fermions (as e.g.~a generation of mirror leptons and/or quarks) or
doublet and triplet fermions (as e.g.~light Higgsinos and gauginos)
remnant from (split) supersymmetry~\cite{Giudice}.  In the next
section we will consider the latter possibility.

\section{\sc Higgsinos and gauginos: the phase transition}
\label{higgsinos}

In this section we will consider the particular case of Higgsinos
($\tilde H_{1,2}$), Winos and Binos ($\tilde W^a,\ \tilde B$) coupled
to the SM Higgs doublet $H$ with the Lagrangian
\begin{eqnarray}
\mathcal L&=&H^\dagger\left(h_2\,\sigma_a\tilde W^a+h^\prime_2\, \tilde
B\right)\tilde H_2 +H^T\epsilon\left(-h_1\,\sigma_a\tilde
W^a+h^\prime_1\, \tilde B\right)\tilde H_1\nonumber\\
&+&\frac{M_2}{2}\,\tilde W^a\tilde W^a
+\frac{M_1}{2}\, \tilde B\tilde B+\mu\, \tilde H^T_2\epsilon \tilde H_1+
h.c.
\label{lagrangiano}
\end{eqnarray}
where $\epsilon=i\sigma_2$ and the Yukawa couplings $h_{1,2}$ and
$h^\prime_{1,2}$ are arbitrary~\footnote{The matching at high scale
with the MSSM couplings would be $h_2=g\sin\beta/\sqrt{2}$,
$h_1=g\cos\beta/\sqrt{2}$, $h_2^\prime=g^\prime\sin\beta/\sqrt{2}$,
$h_1^\prime=g^\prime\cos\beta/\sqrt{2}$, where $g$ and $g^\prime$ are
the $SU(2)$ and $U(1)$ gauge couplings respectively, and the Higgs
doublet is related to the MSSM Higgses by $H=\sin\beta
H_2-\cos\beta\epsilon H^\ast _1$~\cite{Giudice}. The matching with the
couplings of a possible UV completion of the model will be done in
section~\ref{strongc}. For the moment we will just assume that there
is such a UV completion and that it provides the necessary stabilizing
fields as it was discussed in section 2.}.

The chargino mass matrix is
\begin{equation}
\left(
\begin{array}{cc}
M_{2} & h_{1}\,\phi \\
h_{2}\,\phi  & \mu
\end{array}
\right) ,\label{masac}
\end{equation}
and the squared mass matrix has eigenvalues
\begin{equation}
\lambda _{c\pm }=\left( \sqrt{M_{+}^{2}+h_{-}^{2}\phi ^{2}}\pm \sqrt{
M_{-}^{2}+h_{+}^{2}\phi ^{2}}\right) ^{2},  \label{eigenc}
\end{equation}
where $M_{\pm }=\frac{1}{2}\left( M_{2}\pm \mu \right) $, $h_{\pm
}=\frac{1}{ 2}\left( h_{1}\pm h_{2}\right) $.  The mass matrix for
neutralinos is
\begin{equation}
\left(
\begin{array}{cccc}
M_{2} & 0 & -h_{2}\,\phi /\sqrt{2} & h_{1}\,\phi /\sqrt{2} \\
0 & M_{1} & h_{2}^{\prime }\,\phi /\sqrt{2} & -h_{1}^{\prime }\,\phi /\sqrt{2}
\\
-h_{2}\,\phi /\sqrt{2} & h_{2}^{\prime }\,\phi /\sqrt{2} & 0 & -\mu  \\
h_{1}\,\phi /\sqrt{2} & -h_{1}^{\prime }\,\phi /\sqrt{2} & -\mu  & 0
\end{array}
\right) . \label{masan}
\end{equation}
The eigenvalues of this matrix are cumbersome, so we consider the
particular case $M_{1}=M_{2}\equiv M$, $h_{1}=h_{2}\equiv h$, and
$h_{1}^{\prime}=h_{2}^{\prime}\equiv h^{\prime}$. The eigenvalues of
the squared mass matrix are thus
\begin{equation}
\lambda _{n1}=\mu ^{2},\ \lambda _{n2}=M^{2},  \label{eigenn1}
\end{equation}
and
\begin{equation}
\lambda _{n\pm }=\left( M_{+}\pm \sqrt{M_{-}^{2}+ h^2 \phi^2
+2h^{\prime 2}\phi ^{2}}\right) ^{2},  \label{eigenn2}
\end{equation}
In this case the chargino eigenvalues become very similar to those in
Eq.~(\ref{eigenn2}): in particular we have $h_{-}=0$, $h_{+}=h$, and
$M_{2}=M$ in Eq.~(\ref{eigenc}). We can further simplify the problem
by also setting $ h^{\prime }=0$, since in this case
Eqs.~(\ref{eigenc}) and (\ref{eigenn2}) become
\begin{equation}
\lambda _{\pm }=\left( M_{+}\pm \sqrt{M_{-}^{2}+h^{2}\phi ^{2}}\right) ^{2}.
\label{eigen}
\end{equation}
that corresponds to 6 degrees of freedom (a Dirac spinor and a
Majorana spinor) with squared mass $\lambda _{+}$, 6 with squared mass
$\lambda _{-}$, and 2 (light) Majorana particles with masses $\mu$ and
$M$. A total of 16 fermionic degrees of freedom out of which only 12
are coupled to the SM Higgs. Clearly, for $h^{\prime}=0$ the Majorana
particle with mass $M$ is just a pure Bino state of mass $M_1=M$,
which decouples from the other low-energy states and therefore plays
no role in determining the strength of the electroweak phase
transition.  In the following, we shall concentrate on this
particularly simple case.

From Eq.~(\ref{eigen}) it is clear that all the results in this case
will be symmetric under $ \mu \leftrightarrow M$. The simplest
limiting case is when $M_{+}=0$ and $M_{-}=M=-\mu $. In this case the
eigenvalues are degenerate, $\lambda _{\pm }=M^{2}+h^{2}\phi ^{2}$,
with 12 degrees of freedom corresponding to 3 Dirac spinors, and the
situation is identical to the simple example illustrated in the
previous section. One expects that other limits will be less favorable
for the phase transition. For instance if we take $M_{-}=0$
(i.e.~$M_{+}=M=\mu $), the eigenvalues are $\left( M\pm h\phi
\right)^{2}$. This means that, unless $h \phi \geq 2 M$, in the
broken-symmetry phase half of the particles become heavier than in the
symmetric phase but the other half become lighter. Therefore, unless
$M_+$ is also small, less degrees of freedom than in the case $M_+ =
0$ will contribute to this effect.

\begin{figure}[htb]
\psfrag{fit}[][b]{$\phi/T$}\psfrag{vt}[][t]{$\mathcal{F}(\phi,T)/T^4$}
\psfrag{tcp}[][r]{$T'_c$}\psfrag{tc}[][r]{$T_c$} \centering
\epsfysize=4.7cm \leavevmode \epsfbox{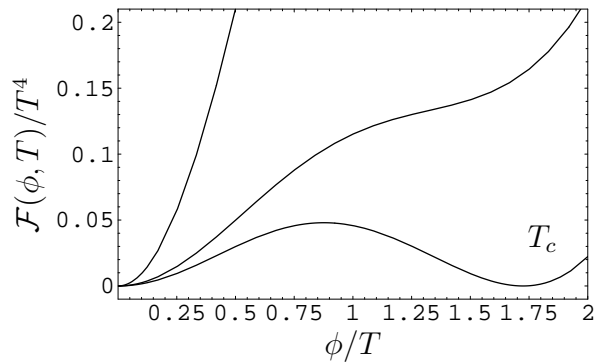}
\caption{\it The free energy at different temperatures for $h=2$,
$m_H=120$~GeV and $\protect\mu =-M\simeq 200$ GeV.}
\label{transi}
\end{figure}

In Fig.~\ref{transi} we plot the free energy at different temperatures
for a Yukawa coupling $h=2$, a Higgs mass of 120 GeV and $\mu=-M\simeq
200$ GeV. We can explicitly see from the figure shape that there is a
first order phase transition with an order parameter $\phi_c/T_c\simeq
1.75$.

\begin{figure}[htb]
\psfrag{xc}[][l]{$\phi_c/T_c$}\psfrag{M}[][b]{$M$ [GeV]} \centering
\epsfysize=6cm \leavevmode \epsfbox{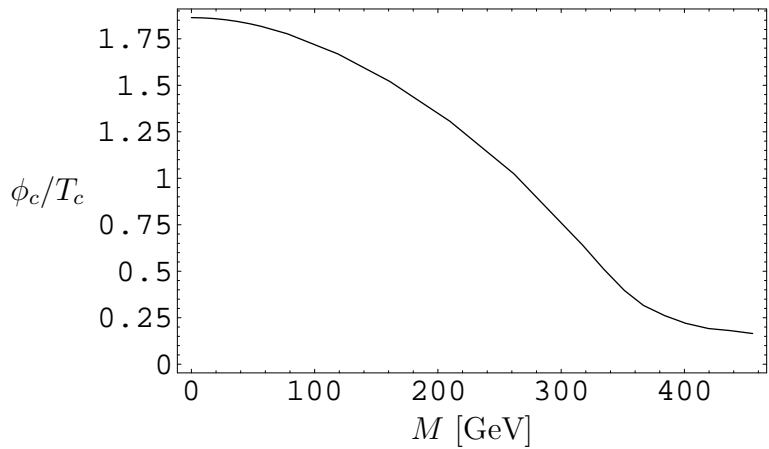}
\caption{\it $\phi_c/T_{c}$ for $m_H=$ 120 GeV and $h=2$ as a function
of $M=-\protect \mu $ in GeV.}
\label{phimantideg}
\end{figure}

In Fig.~\ref{phimantideg} we plot the ratio of the Higgs VEV to the
temperature, evaluated at the critical temperature, as a function of
the mass $M$ for the case $\mu =-M$ and $m_H = 120$~GeV. As expected,
the strength of the phase transition decreases with $M$. This
illustrates the fact that for large $M$ the particle is decoupled
already in the symmetric phase, hence the VEV $\phi_{c}$ has a smaller
value, which corresponds in the $M\to\infty$ limit, to that of the
electroweak phase transition in the Standard Model.

\begin{figure}[htb]
\psfrag{xc}[][l]{$\phi_c/T_c$}\psfrag{M}[][t]{$M$ [GeV]}
\psfrag{mu}[][c]{$\mu$}
\centering
\epsfysize=6cm \leavevmode \epsfbox{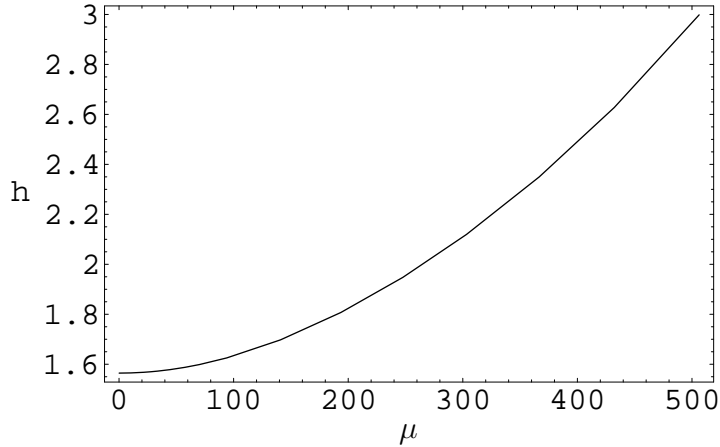}
\caption{\it Yukawa coupling $h$ necessary to get $\phi_c/T_{c}=1$ for
$m_H=$ 120 GeV as a function of $M=-\protect \mu $ in GeV.}
\label{hmu2}
\end{figure}

In Fig.~\ref{hmu2} we plot the values of the Yukawa coupling $h$
necessary to induce a strongly first order phase transition for the
case $M = -\mu$ and $m_H = 120$~GeV. It is clear from the plot that
for such a small value of the Higgs mass, a strengthening of the phase
transition may only be achieved for $h \simgt 1.5$. Let us stress that
this lower bound may be weakened by assuming slighly smaller values of
$\mu_S$. For instance, taking $\mu_S^2$ to be 0.9 times the maximum
value allowed in Eq.~(\ref{bosonstab}) is enough to ensure that values
of $h = 1.5$ and masses $\mu$ of order of 100~GeV are allowed.

We are interested in a strong enough phase transition for
baryogenesis, which means that the Higgs VEV must be large enough for
sphaleron processes to be suppressed in the broken-symmetry phase (in
order to avoid the washout of the BAU after the phase transition). In
the case of Fig.~\ref{phimantideg} this happens up to a value
$M=M_{\max }$ determined by the condition $\phi _{c}/T_{c}\sim 1$.
Observe that the top-quark, with 12 degrees of freedom (similar to the
chargino-neutralino system) and a Yukawa coupling $h_t\simeq 0.7$ in
our normalization, is not able to generate such a strong first-order
phase transition in the SM.

\begin{figure}[htb]
\psfrag{M}[][r]{$M$ [GeV]}\psfrag{MH}[][br]{$m_H$ [GeV]}
\psfrag{h16}[][r]{$h=1.6$}\psfrag{h2}[][r]{$h=2$}
\psfrag{h25}[][r]{$h=2.5$}\psfrag{h3}[][t]{$h=3$}
\centering \epsfysize=6cm \leavevmode \epsfbox{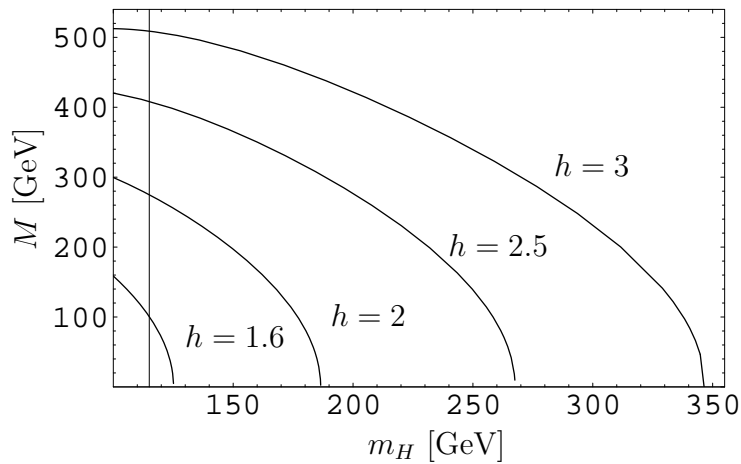}
\caption{\it Contours of $\phi_c/T_c=1$ in the $(M,m_H)$-plane for
$h=1.6,2,2.5,3$ and $M = - \mu$.  The vertical line corresponds to the
experimental lower bound, for a SM-like Higgs, of $m_H=115$ GeV.}
\label{mhiggs}
\end{figure}

It is well known that the value of the Higgs mass plays a prominent
role in the strength of the phase transition in the Standard Model and
extensions thereof. Up to now we have fixed it to a ``minimal'' value,
$m_H=120$ GeV. However our mechanism of strengthening the phase
transition by using strongly coupled fermions, although certainly
sensitive to the value of the Higgs mass, permits to go to higher
values. In Fig.~\ref{mhiggs} we plot, for fixed values of the Yukawa
coupling, the values of $M$ that give $\phi_c/T_c=1$ as a function of
the Higgs mass. We can see that, as expected, there is an upper bound
on the Higgs mass that depends on the Yukawa coupling $m_H^{\rm
max}=m_H^{\rm max}(h)$. Moreover the upper value of the Higgs mass has
an approximate linear behaviour as a function of the Yukawa coupling
$h$ as can be readily deduced from Fig.~\ref{mhiggs}.  Imposing
perturbativity of the theory at the low scale sets a generic upper
limit on the Yukawa coupling $h\simlt \sqrt{4\pi}\sim 3.5$ which from
Fig.~\ref{mhiggs} yields a corresponding upper limit on the Higgs mass
of $m_H\simlt 400$ GeV. In general, for large values of $h$, the
requirement of perturbative consistency of the theory up to high
energies may only be fulfilled by embedding this model into a more
complete theory where couplings become asymptotically free (see
section~\ref{strongc}). For the particularly interesting UV completion
proposed in section~\ref{strongc} the upper limit on the Yukawa
coupling is, as we will see, more restricted, $h \simlt 2$, which
translates into the stronger Higgs mass upper bound $m_H\simlt 175$
GeV for $\mu = -M$ of order 100~GeV.

\begin{figure}[htb]
\psfrag{xc}[][l]{$\phi_c/T_c$}\psfrag{dh}[][br]{$\delta h$}
\centering \epsfysize=6cm \leavevmode \epsfbox{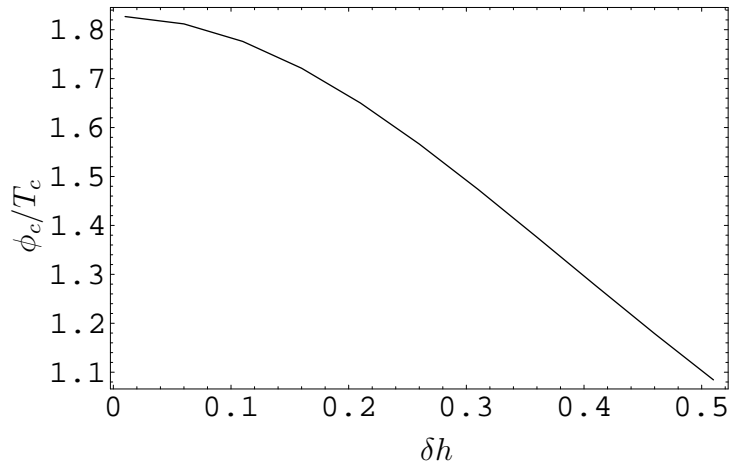}
\caption{\it Values of $\phi_c/T_c$ as a function of $\delta h$, for a
fixed value of the average value $h_{+} = 2$, $m_H=$ 120 GeV and
$M=-\mu=50$ GeV. }
\label{phideltah2}
\end{figure}

The model we have proposed contains Higgsinos $\tilde H_{i}$ and
electroweak gauginos $\tilde W^a$, $\tilde B$, coupled to the SM Higgs
through the Lagrangian (\ref{lagrangiano}) with Yukawa couplings
$h_i$, $h^\prime_i$ and Majorana masses $M_i$.  We have concentrated
on the case $h^\prime_i=0$, in which the Bino state decouples from the
other low energy states. Similar results for the phase transition
would be obtained in any model in which the Bino would decouple from
the low energy theory, for instance if $h^\prime_i \ll h_i$ and $M_1$
is much larger than the weak scale.  Although the Bino is absent at
low energies, there is still a state with mass approximately equal to
$|\mu|$ that is a candidate for Dark Matter. This will be analyzed in
section~\ref{dark} where we will show that a modest splitting between
the Yukawa couplings $h_1$ and $h_2$ will be necessary to accomodate
the observed DM energy density, although a large separation is not
permitted by electroweak precision measurements.  In this way
introducing a small splitting between $h_2$ and $h_1$ is not expected
to modify substantially the previous results in this section. In fact
in the approximation of neglecting terms of order $h_{-}^2$ in the
diagonalization of the mass matrices one can prove that the mass
eigenvalues still correspond to one Majorana spinor with mass $\mu$,
two Dirac spinors with squared masses given in Eq.~(\ref{eigenc}) and
two Majorana spinors with squared masses,
\begin{equation}
\lambda _{\pm }=\left( M_{+}\pm \sqrt{M_{-}^{2}+h_+^2\,\phi ^{2}}\right) ^{2},
\label{eigendelta}
\end{equation}
which of course coincides with the degenerate result of
Eq.~(\ref{eigen}) in the limit of $h_1=h_2$. The result of
Eq.~(\ref{eigendelta}) also proves that the modification of the
strength of the phase transition due to the non-degeneracy will be
small at least in the validity region where Eq.~(\ref{eigendelta})
holds. More explicitly, in Fig.~\ref{phideltah2} we show numerically
the variation of the order parameter $\phi_c/T_c$ away of the
degenerate point as a function of $\delta h = h_{-}$ for $h_+=2$ and
$M=-\mu=50$ GeV. We see that the phase transition is weakened by
$\simlt 20\%$ for $h_{-}/h_+\simlt 0.3$.

\begin{figure}[htb]
\psfrag{xc}[][l]{$\phi_c/T_c$}\psfrag{vf}[][b]{$\varphi$ (rad)} 
\psfrag{pd}[][c]{$\pi/2$}\psfrag{p}[][c]{$\pi$}
\psfrag{tpd}[][c]{$3\pi/2$}\psfrag{dp}[][c]{$2\pi$}
\psfrag{tcp}[][c]{$3\pi/4$}\psfrag{ccp}[][c]{$5\pi/4$}
\centering
\epsfysize=6cm \leavevmode \epsfbox{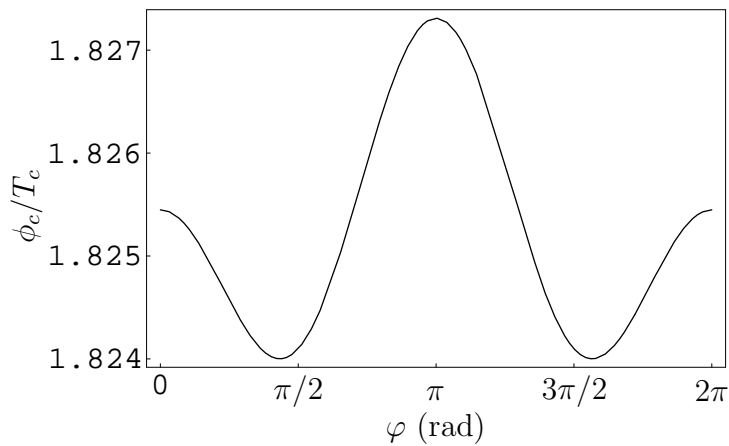}
\caption{\it $\phi_c/T_{c}$ for $h=2$ and $|\mu| =M=50$ GeV as a
function of $\varphi=\arg(\mu)$.}
\label{fivarfi}
\end{figure}

Up to this point we have discussed the electroweak phase transition in
the absence of CP-violating phases. However for the baryogenesis
mechanism to work we need a non-vanishing CP-violating phase in the
parameters of the theory that will trigger baryon number
generation. As in the MSSM studies, we shall consider real Yukawa
couplings (related to gauge couplings in the MSSM) and Majorana
gaugino masses, and a complex mass parameter $\mu = |\mu|
e^{i\varphi}$. To conclude this section we would like to consider how
the phase transition and in particular the order parameter
$\phi_c/T_c$ vary with the phase $\varphi$. The result is presented in
Fig.~\ref{fivarfi}
%
where we plot $\phi_c/T_c$ as a function of $\varphi$ for
$M_2=M_1=|\mu|=50$ GeV and Yukawa couplings $h_+=2$, $h_-=0$.  The
plot at $\varphi=\pi$ is consistent with the corresponding results
that can be read off from Fig.~\ref{phimantideg} while the strength of
the first order phase transition decreases by only a tiny amount for
$\mathcal O(1)$ $CP$ violating phases,~i.e. for $\varphi=\pi/2$.

\section{\sc CP-Violating Sources and Baryogenesis}
\label{cp}

The chargino sector in the model presented in section~\ref{higgsinos}
has a similar structure to the chargino sector in the minimal
supersymmetric Standard Model. The only difference is that the
couplings $g\sin\beta/\sqrt{2}$ and $g\cos\beta/\sqrt{2}$ are replaced
by arbitrary couplings $h_2$ and $h_1$, respectively, as can be seen
in the corresponding mass matrix~(\ref{masac}). As in the MSSM the
CP-violating phase can have its origin, after field redefinitions, in
the phase $\varphi$ of the (complex) $\mu$-parameter. A general method
for computing the effects of CP-violating mass terms on particle
distributions was introduced in Ref.~\cite{HN} leading to an efficient
transport of CP-violating quantum numbers into the symmetric phase
where weak sphalerons are active and can trigger electroweak
baryogenesis for all bubble wall widths. The method was adapted to the
MSSM by a number of papers~\cite{CQRVW}-\cite{CQSW} where a set of
coupled differential equations, that include the effect of diffusion,
particle number changing reactions and CP violating terms, were solved
to find various particle number densities diffused from the bubble
wall, where CP-violation takes place, to the symmetric phase where
sphalerons are active. These methods can be adapted to the present
model by just considering the particular structure of the chargino
mass matrix given by Eq.~(\ref{masac}). We will further make the
simplifying assumption that all $CP$ violation resides in the
fermionic sector. Otherwise there should be extra contributions to the
$CP$ violating currents from the bosonic (stabilizing) fields,
although these contributions are expected to be suppressed with
respect to the fermionic ones because the stabilizing fields are
heavier than the fermions.  So from this point of view our results
will be mostly conservative.

In particular we will follow the formalism of Refs.~\cite{CMQSW,CQSW}
where a method was developed to compute the CP-violating sources
induced by the passage of a bubble wall in a system of fermions that
interact in a way similar to the one described above, in an expansion
of derivatives of the Higgs fields. The method allows for the
computation of the currents in a resummation to all orders of the
Higgs vacuum expectation value effects. It was found that there are
two different CP-violating sources from the chargino sector which the
total baryon asymmetry depends upon. The leading contribution is
provided by
\begin{equation}
\epsilon_{ij}H_i\partial_\mu H_j=v^2(T)\partial_\mu\beta
\label{source1}
\end{equation}
that is proportional to the variation of the angle
$\beta=\arctan\left[v_2(T)/v_1(T)\right]$ at the wall of the expanding
bubble. The source (\ref{source1}) has a resonant behaviour for
$M_2=|\mu|$ and it is the leading contribution in the MSSM. However,
the Higgs sector of our model (which contains just the SM Higgs
doublet) can be considered as the $m_A\to\infty$ limit of the MSSM
Higgs sector (where $m_A$ is the pseudoscalar mass) in which case
$\partial_\mu \beta\to 0$ and the source (\ref{source1}) does not
contribute to the diffusion equations.

The second contribution to the baryon asymmetry is proportional to
\begin{equation}
H_1\partial_\mu H_2+H_2\partial_\mu
H_1=v^2\cos(2\beta)\partial_\mu\beta+v\partial_\mu v\sin(2\beta).
\label{source2}
\end{equation}
In the limit $m_A\to\infty$ only the second term in (\ref{source2})
survives. Moreover we will consider in this section $h_1\simeq
h_2\equiv h$ (i.e.~$\tan\beta\simeq 1$)~\footnote{This choice will be
motivated by the contribution of charginos and neutralinos to the
electroweak parameter $T$ as we will see in section~\ref{precision}.}
in which case the only remaining source is proportional to
$\partial_\mu v^2$. This region is, as it was proven in
Ref.~\cite{CQSW}, very insensitive to the resonance region relating
$M_2$ and $|\mu|$ and it provides a very natural region of parameters
where electroweak baryogenesis can hold. Although in the MSSM such a
region provided a very tiny amount of baryon asymmetry, in the present
model all effects are enhanced by the strong Yukawa couplings of the
Higgs to charginos as we will see in this section. In the MSSM there
is an additional suppression of the source, Eq. (\ref{source2}), due
to the large values of $\tan\beta$ necessary to fulfill the LEP bounds
on the lightest CP-even Higgs boson mass~\cite{Barate:2003sz},
$\tan\beta$ of order 10. As stressed above in this model, instead,
$\tan\beta \simeq 1$.

Following the formalism of Refs.~\cite{CMQSW,CQSW} the solution of the
diffusion equations, in the limit where the strong sphaleron
($\Gamma_{ss}$) and Yukawa processes ($\Gamma_Y$) are fast enough,
provide quark number density for third generation doublets $n_Q$ and
singlets $n_T$ as functions of the number density for the Higgs
doublet coupled to the top quark $n_H$, as
\begin{eqnarray}
n_Q(z)&=&\frac{k_Q(9k_T-k_B)}{k_H(k_B+9k_Q+9k_T)}\ n_H(z)
+\mathcal{O}\left(\frac{1}{\Gamma_{ss}},\frac{1}{\Gamma_Y}\right)\nonumber\\
n_T(z)&=& -\frac{k_T(9k_Q+2k_B)}{k_H(k_B+9k_Q+9k_T)}\
n_H(z)+\mathcal{O}\left(\frac{1}{\Gamma_{ss}},\frac{1}{\Gamma_Y}\right)
\label{densi}
\end{eqnarray}
where $k_i$ are statistical factors~\cite{GS}
\begin{eqnarray}
k_B(m^2)&=& \frac{3}{2\pi^2}\,\int_0^\infty
dp\, \frac{p^2}{\sinh^2\left(\sqrt{(p^2+m^2)/T^2}/2\right)}
\label{kB}\\
k_F(m^2)&=& \frac{3}{2\pi^2}\,\int_0^\infty
dp\, \frac{p^2}{\cosh^2\left(\sqrt{(p^2+m^2)/T^2}/2\right)}
\label{kF}
\end{eqnarray}
that satisfy the condition $k_F(0)=1$ ($k_B(0)=2$) for Weyl fermions
(complex bosons). In turn the density $n_H(z)$~\footnote{The spatial
coordinate $z$ is transverse to the bubble wall and we are neglecting
the bubble curvature.} is obtained from the diffusion equations as a
function of the particle number changing rates, CP-violating sources
and diffusion constants, as explained in Ref.~\cite{CQSW}, yielding
from Eq.~(\ref{densi}) the quark number densities $n_{Q,T}$.

In order to evaluate the baryon asymmetry generated in the broken
phase $n_B$ we first need to compute the density of left-handed quarks
and leptons, $n_L$, in front of the bubble wall in the symmetric
phase. These chiral densities bias weak sphalerons to produce a net
baryon number. Considering particle species that participate in fast
particle number changing transitions, and neglecting all Yukawa
couplings except those corresponding to the top quark, only quark
doublets do contribute to $n_L$. Then assuming that all quarks have
nearly the same diffusion constant it turns out that~\cite{HN}
$n_{Q_1}=n_{Q_2}=2(n_Q+n_T)$ and therefore from Eq.~(\ref{densi})
\begin{eqnarray}
n_L(z)&=&5n_Q(z)+4n_T(z)=A\,
n_H(z)
+\mathcal{O}\left(\frac{1}{\Gamma_{ss}},\frac{1}{\Gamma_Y}\right)\nonumber\\
A&=& \frac{5\,k_Qk_B+8\, k_T k_B-9\,k_Q k_T}{k_H(k_B+9\,k_Q+9\,k_T)}\ .
\label{densidad}
\end{eqnarray}

It turns out that the baryon asymmetry can be written as~\cite{CQSW}
\begin{equation}
n_B=-\frac{n_F\Gamma_{ws}}{v_\omega}\int_{-\infty}^0 dz\ n_L(z)
\exp\left(-\frac{5 n_F \Gamma_{ws} z}{4 v_w} \right)
\label{nb}
\end{equation}
where $n_F=3$ is the number of families, $v_\omega$ the bubble wall
velocity and $\Gamma_{ws}=6\kappa \alpha_w^5 T$, where $\kappa\simeq
20$~\cite{sphalT}, is the weak sphaleron rate.

For the model presented in this paper, where squarks and the non-SM
Higgs bosons are superheavy, $m_Q,\,m_T,\,m_B\gg T_c\simeq 100$ GeV
($k_B=k_T\simeq 3$, $k_Q\simeq 6$ and $k_H\simeq 8$), it turns out
that the coefficient in Eq.~(\ref{densidad}) is $A\simeq 0$. This SM
suppression was already pointed out by Giudice and
Shaposhnikov~\cite{GS} and consequently the baryon asymmetry $n_B$ in
our model is produced by sub-leading effects. Assuming
$\Gamma_Y\gg\Gamma_{ss}$ we can go beyond the approximation of
Eq.~(\ref{densi}) and work out corrections of $\mathcal
O(1/\Gamma_{ss})$. This was done in Ref.~\cite{HN} leading to an
$\mathcal O(1/\Gamma_{ss})$ correction to $n_L(z)$, $\Delta_{ss}
n_L(z)$, in our model as
\begin{equation}
\Delta_{ss} n_L(z)=-\frac{3}{112}\frac{D_q
n_H^{\prime\prime}(z)-v_\omega n_H^{\prime}(z)}{\Gamma_{ss}}
\label{deltass}
\end{equation}
where $D_q\simeq 6/T$ is the quark diffusion constant and the strong
sphaleron rate is given by
$\Gamma_{ss}=6\kappa^\prime\frac{8}{3}\alpha_s^4T$, where
$\kappa^\prime$ is an order one parameter~\cite{HN}. When
(\ref{deltass}) is inserted into (\ref{nb}) it produces the baryon
asymmetry generated by the sub-leading $\mathcal O(1/\Gamma_{ss})$
effects. We have numerically checked that this correction is
insufficient to generate the observed baryon asymmetry of the
universe.

\begin{figure}[htb]
\vspace{1cm}
\centering \epsfysize=7cm \leavevmode \epsfbox{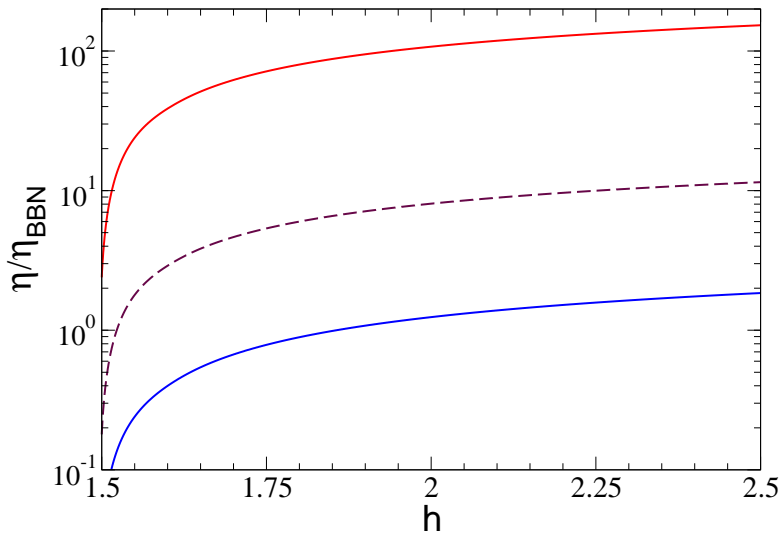}
\caption{\it The ratio $\eta/\eta_{\rm BBN}$ as a function of the
Yukawa coupling $h$ for $\mu = - M_2 \exp(i\varphi)$, $M_2=50$ GeV,
maximal CP-violating phase, $\sin\varphi = 1$, and bubble parameters
$L_\omega=10/T_c$, $v_\omega=0.1$. Left-handed squarks and
right-handed sbottoms are heavy (in the few TeV range). The lower
(upper) solid line corresponds to heavy (light) right-handed stops,
$m_T\simgt 1$ TeV ($m_T\simeq 100$ GeV). Dashed line corresponds to
right-handed stops with $m_T\simeq 500$ GeV.}
\label{ewbg2}
\end{figure}

Another, more important, correction that can lead to a non-zero value
of the baryon asymmetry to leading order in $\Gamma_{ss}$ are Yukawa
and gauge radiative corrections to statistical coefficients $k_i$ (or
equivalently thermal masses)~\cite{GS}. Expanding Eq.~(\ref{kF}) in a
power series of $m^2/T^2$ we can write
\begin{equation}\label{kFexp}
k_F(m^2)=1-\frac{3}{2\pi^2}\,\frac{m^2}{T^2}+\mathcal O(m^4/T^4)\ .
\end{equation}
Keeping only the strong gauge ($g_s$) and top Yukawa ($h_t$) couplings
one obtains in our model the statistical coefficients,
\begin{eqnarray}
k_T&=&3(1+\Delta_s+\Delta_Y)\nonumber\\
k_B&=&3(1+\Delta_s)\nonumber\\
k_Q&=&6(1+\Delta_s+\frac{1}{2}\Delta_Y)
\end{eqnarray}
with
\begin{equation}
\Delta_s=-\frac{g_s^2}{2\pi^2},\quad \Delta_Y=-\frac{3 h_t^2}{8\pi^2}
\end{equation}
where $h_t\simeq 1/\sqrt{2}$ is the top quark Yukawa coupling and
$g_s\simeq 1.2$ the strong gauge coupling. Numerically,
$|\Delta_s|\sim 7\times 10^{-2}$ is more important than
$|\Delta_Y|\sim 2\times 10^{-2}$ but since the strong correction to
all quarks is universal it cancels in the contribution to the baryon
asymmetry. In fact to linear order in $\Delta_i$ one can write the
density $n_L(z)$ in Eq.~(\ref{densidad}) as
\begin{equation}
n_L(z)=-\frac{3}{16}\,\Delta_Y\ n_H(z)
\end{equation}
and the proportionality coefficient turns out to be $A\sim 4\times
10^{-3}$. This is the reduction factor we get from such a sub-leading
effect. The numerical calculation of the baryon-to-entropy ratio
$\eta$ is presented in Fig.~\ref{ewbg2} (lower solid line) where we
%
plot its ratio to the experimentally determined value $\eta_{\rm
BBN}=(8.7\pm 0.3)\times 10^{-11}$~\cite{PDG} as a function of the
Yukawa coupling $h$ and we have fixed the CP-violating phase
$\sin\varphi=1$. In fact the phase $\sin\varphi$ that would be
required for fixing $\eta=\eta_{\rm BBN}$ is given by the inverse
value plotted in Fig.~\ref{ewbg2}.  We have chosen the case $\mu =
-M_2 \exp(i\varphi)$, and small values of $M_2 = |\mu| \simeq 50$ GeV,
where the phase transition is favoured, and typical values of the
bubble width and velocity~\footnote{A general feature of first-order
phase transitions is that the release of latent heat causes a
slow-down of bubble expansion~\cite{ariel}. The electroweak
bubble-wall velocity thus decreases during the phase
transition~\cite{h95} from its initial value $v_w \sim
10^{-1}$~\cite{js01} given by the friction of the plasma.  Calculating
the exact value of $v_w$ is out of the scope of this paper. However,
as noticed in Ref.~\cite{am04}, the effect of the velocity variation
on the BAU is likely to be an $\mathcal O(1)$ one effect and should
not modify the main conclusions of this paper.}.  Since the
computation of the baryon asymmetry has been performed by ignoring
corrections of order one, the main conclusion one can extract from the
results of Fig.~\ref{ewbg2} is that CP-violating phases such that
$\sin\varphi$ is of order one are necessary to obtain a value of the
baryon asymmetry consistent with the experimentally determined values,
for any value of $h \simgt 1.5$.

The amount of generated baryon number density can be increased if some
squark is light enough to be in equilibrium with the thermal bath
during the phase transition, in which case the SM suppression is
avoided.  The typical case that was considered in previous studies is
that of a light right-handed stop~\cite{CQW} that corresponds to
values of the supersymmetry breaking soft masses $m_Q,\, m_B\gg T_c$
and $m_T\simlt T_c$. In that case the statistical coefficients are
given by $k_Q\simeq 6$, $k_T\simeq 9$, $k_B\simeq 3$ and $k_H\simeq 8$
and the coefficient $A$ in Eq.~(\ref{densidad}) does not vanish to
leading order in $\mathcal O(1/\Gamma_{ss})$. In fact it is given by
$A\simeq 1/6$ and this produces an enhancement factor of $\mathcal
O(100)$ with respect to the case where all squarks are
superheavy. This enhancement factor produces larger values of $\eta$
(and so smaller phases are allowed) as can be seen in Fig.~\ref{ewbg2}
upper solid line. Now fixing $\eta=\eta_{\rm BBN}$ can be consistent
with phases $\sin\varphi\simeq 10^{-2}$. Notice finally that a similar
enhancement would also appear if other squark species
(i.e.~right-handed sbottom or left-handed doublet) are light; this
effect is not particularly linked to the lightness of right-handed
stops.

As mentioned above, the upper solid line in Fig.~\ref{ewbg2}
corresponds to the extreme case where there are no extra bosons in the
low energy spectrum, or equivalently stop masses in the TeV range or
larger~\footnote{For instance for $m_T=10\, T_c\simeq 1$ TeV the stop
contribution to $k_T$ is $\sim 3\times 10^{-3}$ and $A\sim 1.8\times
10^{-4}$, much smaller than the previously considered thermal
effects.}.  Of course there can be intermediate situations where the
stop is heavy but still does not fully decouple from the thermal
plasma.  In this case it contributes to the statistical factor $k_T$
with some small value that can significantly contribute to the
$A$-parameter and departure its value from zero. For instance if
$m_T=5\, T_c\simeq 500$ GeV, its contribution to the statistical
factor $k_T$ as given from (\ref{kB}) is $\sim 0.24$ which produces in
(\ref{densidad}) a value $A\sim 1.2\times 10^{-2}$ and enhances the
value of $\eta$ from its value with fully decoupled right-handed
stops. The corresponding value of $\eta$ is plotted in
Fig.~\ref{ewbg2} (dashed line). We can see that the CP-violating phase
for $\eta=\eta_{\rm BBN}$ is now $\sin\varphi\simeq 0.1$. A similar
effect would be produced by other not-so-heavy third generation
squarks.

\section{\sc Electron Electric Dipole Moment}
\label{edm}

In the absence of light squarks, baryon number generation demands the
presence of large phases in the chargino and neutralino sectors. In
the previous section, we assumed gaugino masses and Yukawa couplings
to be real, and therefore the relevant phase is the one of the $\mu$
parameter. Although one-loop corrections to the electron electric
dipole moments are suppressed in the limit of heavy fermions, as has
been stressed in Ref.~\cite{Apostolos}, two-loop contributions become
relevant. In this section we will evaluate the two-loop contribution
to the electron electric dipole moment from the fermion and Higgs
sector~\footnote{We will assume here that squarks and stabilizing
bosons are heavy enough not to contribute appreciably to electric
dipole moments.  However when particular UV completions of this model
will be considered this assumption should be re-checked and if the new
contributions are relevant they should be added to the fermionic
ones.}.

\begin{figure}[bht]
\psfrag{si}[][c]{$\sin(\varphi) = 1$} \centering
\centering \epsfysize=7cm \leavevmode \epsfbox{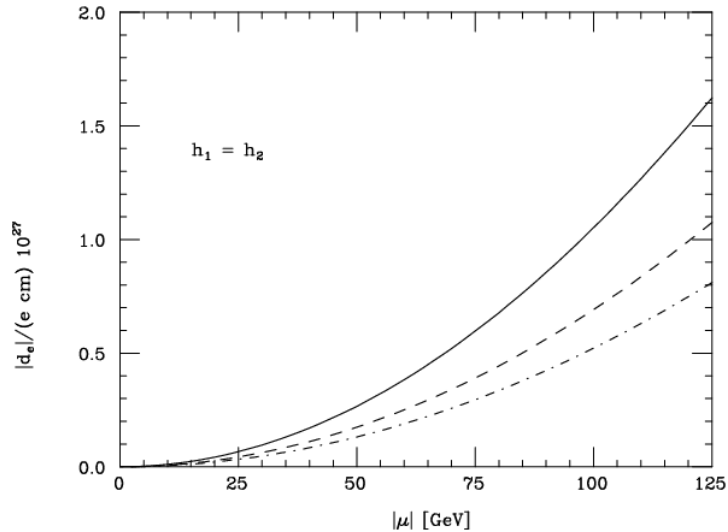}
\caption{\it Electric dipole moment of the electron as a function of
$|\mu|$ for maximal CP-violating phase, $\sin\varphi = 1$, and $h_{-}
= 0$. The Higgs boson mass and Yukawa couplings are fixed to $m_H =
120$~GeV, $h_{+} = 1.5$ (solid line); $m_H = 120$~GeV, $h_{+} = 2$
(dashed line); and $m_H = 200$~GeV, $h_{+} = 2$ (dot-dashed line).}
\label{edmfig}
\end{figure}

In Fig.~\ref{edmfig} we plot the chargino contribution to the electron
electric dipole moment~\cite{Chang}, for $\mu = -M_2 \exp(i\varphi)$,
with $M_2$ real and positive ($M_2 = |\mu|$), maximal CP-violating
phase, $\sin\varphi = 1$, $h_{-} = 0$, and for $h_{+} = 1.5$, $m_H =
120$~GeV (solid line); $h_{+} = 2$, $m_H = 120$~GeV (dashed line); and
$h_{+} = 2$, $m_H = 200$~GeV (dot-dashed line). We have verified that
the results vary only slightly for non-vanishing values of
$|h_{-}/h_{+}| \simlt 0.1$, which, as we will show in
sections~\ref{dark} and~\ref{precision}, are preferred by dark matter
and precision electroweak constraints.
As discussed above, for values of $h_{+} < 1.5$, the electroweak phase
transition is weakly first order and the generated baryon asymmetry is
not preserved. For slightly larger values, $h_{+} \simlt 1.6$, small
values of $|\mu| < 100$~GeV and Higgs mass values smaller than 125~GeV
are demanded in order to make the phase transition strongly first
order. The present electric dipole moment bound, $|d_e|/(e cm) < 1.7
\; 10^{-27}$~\cite{Regan:2002ta} does not put a bound on this model
for these values of $h_{+}$ and $|\mu|$. For larger values of $h_{+}$
and similar values of $|\mu|$, Fig.~\ref{edmfig} shows that the
electron electric dipole moment contributions become smaller and in
addition, as shown in Fig.~\ref{ewbg2}, smaller phases are demanded
for the baryon number generation. Therefore, the electric dipole
moment bounds become even weaker in this case.

However the anticipated improvement of a few orders of magnitude in
this quantity~\cite{Kawall:2004nv} will be sufficient to test this
model even for values of $h_{+} = 2$, provided the values of $|\mu|$
are in the range necessary to obtain a good dark matter source and
avoid the LEP invisible constraints: For large values of $M_1$, (or
for $M_2 = M_1$) and small values of $h_{-}$, the lightest neutralino
mass is mainly a Higgsino with mass approximately equal to
$|\mu|$. For $h_{+} \geq 1.5$, the chargino and the second lightest
neutralino are heavier than 200~GeV and therefore, the stronger
experimental bound comes from the $Z$ invisible width.  Surprisingly,
for $h_{-} = 0$, the tree-level coupling of the lightest neutralino to
the $Z$ vanishes and therefore the invisible width bounds become very
weak. However as we shall analyze in section~\ref{dark}, assuming
thermal production of the lightest neutralino, an acceptable relic
density may only be obtained for a small but non-vanishing coupling to
the $Z$.  Quite generally, if the dark matter density is determined by
the s-channel $Z$-annihilation cross section of a neutralino, an
acceptable relic density may only be obtained for values of the
neutralino mass larger than 35 GeV, what in this case implies $|\mu|
\simgt 35$~GeV~\cite{Menon:2004wv}. For such a range of values of
$|\mu|$, a bound on the electric dipole moment of about $10^{-29}$ e
cm will be enough to test this model for $h_{+} \simlt 2$ and
$\sin\varphi \simeq {\cal{O}}(1)$, as required for baryogenesis in the
absence of light squarks.

\section{\sc Dark Matter}
\label{dark}

One of the most attractive features of the model presented above is
that the particles that lead to a strengthening of the electroweak
phase transition are the same as the ones leading to a generation of
the baryon number at the weak scale. It would be most important if the
same particles would also provide a good Dark Matter candidate.  As
stated in the introduction stable, neutral, weakly interacting
particles lead naturally to a Dark Matter relic density of the order
of the one present in nature. As we will see, under the assumption of
an $R$-Parity symmetry, the lightest neutralino of the model presented
above becomes a good Dark Matter candidate.

Let us work in the simplest case, in which the Bino mass $M_1$ takes
large values ($M_1\gg M_2$) and mixes only weakly with the Higgsino
($h'_{1,2}\ll h_{1,2}$). In such a case, due to the strong Yukawa
couplings $h_1$ and $h_2$, the charginos and two of the neutralinos
acquire masses of about $h_+ v$. The mass of the lightest neutralino
is close to $|\mu|$ and the lightest neutralino is therefore an almost
pure Higgsino state.

Assuming that all squarks, stabilizing bosons and heavy Higgses, if
present, are considerably heavier than the lightest Higgsino, the
states which determine the neutralino annihilation cross section are
the light SM-like Higgs boson and the $Z$-gauge boson. Due to the
small coupling of the SM Higgs boson to quarks and leptons, the
annihilation cross section via s-channel Higgs boson production is
very small, unless the neutralino mass is very close to $m_h/2$, a
quite unnatural possibility that we shall discard for the aim of this
work.  Hence the annihilation cross section is governed by the
coupling of the lightest neutralino to the $Z$-gauge boson.

The coupling of a neutralino state to the $Z$-gauge boson is
proportional to the difference of the square of the components
$N_{\tilde{\chi}i}$ of the neutralino into the two weak Higgsino
states $\tilde{H}_i$,
\beq
g_{\tilde{\chi}Z} \propto \left(|N_{\tilde{\chi}1}|^2 -
|N_{\tilde{\chi}2}|^2 \right)
\eeq
This difference vanishes for values of $h_1 = h_2$, and increases for
increasing values of $h_{-}$. Considering small differences between
the values of $h_1$ and $h_2$, the lightest neutralino, with mass
approximately equal to $|\mu|$, is given by
\beq
\tilde{\chi} \simeq \frac{ h_1}{\sqrt{h_1^2 + h_2^2}}\, \tilde{H}_2
+ \frac{ h_2}{\sqrt{h_1^2 + h_2^2}}\, \tilde{H}_1
\eeq
and hence
\beq
g_{\tilde{\chi}Z} \propto \frac{h_2^2 - h_1^2}{h_1^2 + h_2^2}\ .
\label{coupling}
\eeq
The annihilation cross section is proportional to the square of
$g_{\tilde{\chi}Z}$ and inversely proportional to the square of the
difference between the lightest neutralino mass and the resonant mass
value, $M_Z/2$. Therefore, the smaller the coupling
$g_{\tilde{\chi}Z}$, the closer the modulus of the parameter $\mu$
should be to the resonant mass value. Hence, in order to get a value
of the relic density consistent with the one determined by WMAP, there
must be a correlation between the departure of $|\mu|$ from $M_Z/2$
and the difference of the Yukawa couplings of the two Higgsinos to the
Wino and Higgs field.

The numerical estimates of the values of $|\mu|$ for a given value of
$h_{+}$ have been obtained by computing the relic density, which is
inversely proportional to the thermal average annihilation cross
section,
\beq \Omega\,h^2 =
\frac{(1.07\times10^9\,\mbox{GeV}^{-1})}{M_{Pl}}
\left(\int_{x_f}^{\infty}dx\,\frac{\left<\sigma\,v\right>(x)}{x^2}
g_{*}^{1/2}\right)^{-1}, \eeq
where $x = m_{\tilde{\chi}}/T$, $m_{\tilde{\chi}}$ is the mass of the
lightest neutralino particle and $T$ the temperature of the
Universe~\cite{Gondolo:1990dk}.  The value of the variable $x$ at the
freeze-out temperature, $x_f = m_{\tilde{\chi}}/T_f$, is given by the
solution to the Eq.~\cite{Kolb:vq}
\beq x_f =
\ln\left[\frac{0.038\,(g/g_{*_f}^{1/2})\,m_{\tilde{\chi}}\,M_{Pl}
\left<\sigma\,v\right>(x_f)}{x_f^{1/2}}\right], \eeq
where $g=2$ is the number of degrees of freedom of the neutralino,
$g_{*_f}$ is the total number of relativistic degrees of freedom at
temperature $T_f\;$, and $M_{Pl}$ is the Planck mass.  The thermal
average of the annihilation cross section may be computed by standard
methods and, for a particle of mass $m_{\tilde{\chi}}$, is given
by~\cite{Gondolo:1990dk}
\beq \left<\sigma\,v\right> =
\int_{4m_{\tilde{\chi}}^2}^{\infty}ds\, \sqrt{s-4m_{\tilde{\chi}}^2}\,
W\,K_1({\sqrt{s}}/{T})\Big/16\,
m_{\tilde{\chi}}^4\,T\,K_2({m_{\tilde{\chi}}}/{T}),
\label{sigv}
\eeq
where $s$ is the usual Mandelstam parameter, $K_1$ and $K_2$ are
modified Bessel functions, and the quantity W is defined to be
\beq
W = \int\left[\prod_f\frac{d^3p_f}{(2\pi)^3\,2E_f}\right]\;
(2\pi)^4\delta^{(4)}(p_1+p_2-\sum_fp_f)\;|\mathcal{M}|^2,
\eeq
where $|\mathcal{M}|^2$ is the squared matrix element averaged over
initial states, and summed over final states.
\vspace{1cm}
\begin{figure}[htb]
\centering \epsfysize=7cm
\leavevmode \epsfbox{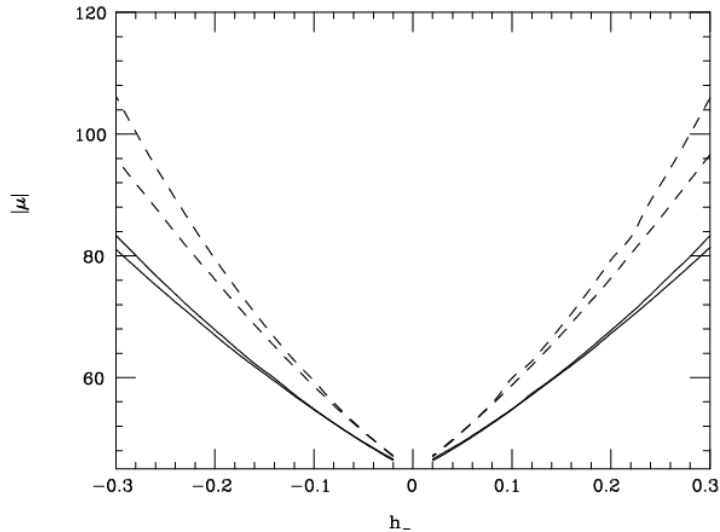}
\caption{\it Values of the $\mu$ parameter leading to a value of the
neutralino relic density consistent with the central value of the WMAP
observation, as a function of $h_{-}$, for $\mu = - M_2
\exp(i\varphi)$ and $M_2 = |\mu|$, assuming that the average value of
the Yukawa couplings $h_{+}$ is fixed to the value $h_{+}$ = 2 (solid
lines) and $h_{+} =$ 1.5 (dashed lines). Lower and upper lines for
each set of $h_{+}$ values correspond to $\varphi = 0$ and $\varphi =
\pi/2$, respectively.}
\label{relic}
\end{figure}

In Fig.~\ref{relic} we plot the required value of $|\mu|$ in order to
obtain the central value of the relic density consistent with the
recent experimental results $\Omega h^2\simeq \Omega_{\rm
WMAP}h^2=0.11\pm 0.01$~\cite{PDG} as a function of $h_{-}$.  The
average values of the Yukawa couplings have been fixed to $h_{+} =1.5$
(dashed lines) and $h_{+} = 2$ (solid lines).  Lower and upper lines
for each set of $h_{+}$ values correspond to two different values of
the CP-violating phase, $\varphi = 0$ and $\varphi = \pi/2$
respectively.  Larger values of $|h_{-}|$ lead to larger values of
$|\mu|$.  We show the results only for small values of $|h_{-}|$,
since as we will discuss in the next section, large values of
$|h_{-}|$ lead to unacceptably large corrections to the precision
electroweak observables.  A non-vanishing phase has only a significant
impact on the relic density determination for values of $h_{+} \simeq
1.5$ and $|\mu| > 80$~GeV ($|h_{-}| \simgt 0.2$), for which the phase
transition is no longer strongly first order and/or unacceptably large
corrections to the precision electroweak data are
generated. Otherwise, the variation induced by the CP-violating phases
is of the order of (or smaller than) the present experimental
uncertainty in the relic density. Indeed, a one sigma variation of the
relic density results for values of $|\mu| \simeq 80$~GeV (60~GeV)
would be obtained by varying $|\mu|$ by about 3~GeV (1.5~GeV).

It is important to compare the results shown in Fig.~\ref{relic} with
those in Fig.~\ref{mhiggs}.  For $h_{+} \simlt 1.6$ it follows from
Fig.~\ref{mhiggs} that in order to preserve a strongly first order
phase transition one needs $|\mu| < 100$ GeV and a Higgs mass smaller
than 125~GeV.  As seen in Fig.~\ref{relic}, such small values of
$|\mu|$ are also consistent with Dark Matter constraints, provided
that $|h_{-}| \simlt 0.3$.  As emphasized before, and as we will show
in detail in section 7, the requirement of consistency with the
precision electroweak data further constraints the allowed parameter
space.

For $h_{+} = 2$, the constraints coming from the requirement of
a sufficiently strong phase transition are much weaker.  Values of
$|\mu|$ and of the Higgs mass as large as 200~GeV are consistent with
this constraint. This is a much larger region of values of the
parameter $|\mu|$ than the one consistent with Dark Matter
constraints, for the small values of $|h_{-}|$ that are required by
precision electroweak data (see section 7). Therefore, also in this
case the allowed region of parameters may only be determined once the
analysis of the constraints coming from the consistency with
precision electroweak data are evaluated. 

Finally notice that, for a fixed value of $h_+$ in Fig.~\ref{relic},
the region above (below) the line $\Omega=\Omega_{\rm WMAP}$
corresponds to $\Omega>\Omega_{\rm WMAP}$ ($\Omega<\Omega_{\rm
WMAP}$). Therefore while the region above the corresponding curve is
excluded since it would predict too much DM density, the region below
it is not excluded provided there is another candidate for Dark Matter
in the theory.

\section{\sc Electroweak precision measurements}
\label{precision}

Heavy fermions, with large couplings to the Higgs field, may induce
large corrections to the electroweak precision measurement parameters.
Extra contributions may come from the stabilizing fields, but in this
section we shall assume that the UV completion of the model is such
that they are small. A very trivial (ad hoc) way of achieving this is
if the stabilizing fields are a set of scalar singlet fields strongly
coupled to the Higgs boson. For instance we can consider the case of
$N$ scalar (complex) singlets giving a contribution to the scalar
effective potential given by
\begin{equation}
V_S = h^2|\vec S|^2
|H|^2 + \mu_S^2|\vec S|^2 + \lambda_S |\vec S|^4,
\end{equation}
with $\mu_S^2, \lambda_S > 0$. This set of singlet fields
contributes to the one-loop effective potential 
of the Higgs field as $g_b=2N$
bosonic degrees of freedom with a mass squared
$m^2(\phi)=\mu_S^2+h^2\phi^2$ that corresponds to the typical case of
stabilizing fields introduced in section~\ref{phase}~\footnote{
Strictly speaking, for the stabilization of the
effective potential, it is not necessary 
that the bosonic and fermionic number of degrees
of freedom, $g_b$ and $g_f$, are equal and/or that the corresponding
Yukawa couplings $h_b=h_f$,
as we have been
assuming in section~\ref{phase}. 
It is easy to see that a necessary
stabilization condition is provided by $ g_b h^4_b\geq g_f h^4_f
$.}. As it is obvious this system of stabilizing (singlet) fields
would not contribute to the electroweak precision observables or
to the CP-violating observables analysed in the previous sections. 
In the following, we will just concentrate on the contribution to the
electroweak precision measurement parameters from the fermionic sector
of the theory.

It is well known that if the fermion masses proceed from the usual
contraction of a left-handed fermion doublet and a right-handed
fermion singlet with the Higgs doublet, and if the mass difference of
the fermion components of the doublet field is small, the contribution
of heavy fermions to the $S$-parameter is about $1/6\pi$. On the other
hand the contribution to the $T$-parameter would depend on the size of
the mass difference between the up and down fermions.  Cancellation of
anomalies requires the presence of at least two such new heavy fermion
doublets and therefore the contribution to the $S$-parameter tends to
be large, $S>0.1$.

In the case under analysis, however, the symmetry breaking masses
proceed from the coupling of the Higgsino doublets to the Higgs field
and an $SU(2)_L$-triplet of Winos. Contrary to the standard case of
heavy fermions, the contribution to the $S$-parameter becomes small in
this case. For a given value of the average Yukawa coupling $h_{+}$,
the contribution to the $T$ parameter, instead, becomes sizeable for
large values of $h_{-}$, while the contribution to the parameter $U$
is an order of magnitude smaller than the contribution to $T$.

\begin{figure}[htb]\vspace{1.5cm}
\centering \epsfysize=8cm
\leavevmode \epsfbox{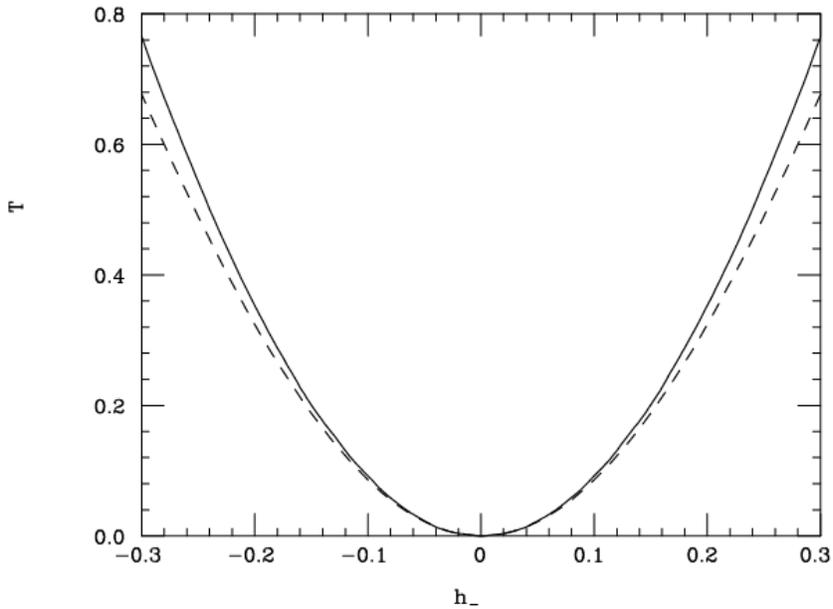}
\caption{\it Chargino and neutralino contributions to the
$T$-parameter as a function of $h_{-}$, assuming that the average
values of the Yukawa couplings has been fixed to $h_{+} = 2$ (solid
line) and $h_{+} = 1.5$ (dashed line) and considering $M = - \mu$
determined as a function of $h_{-}$ from Figure~\ref{relic}.  }
\label{rho}
\end{figure}

We shall work in the limit in which $M_1$ is large. Thus the mixing of
Binos with the Winos and Higgsinos becomes small and therefore the
Binos decouple from the precision measurement analysis.  For large
values of the Yukawa couplings, the lightest neutralino has a mass
close to $|\mu|$ and a coupling to the $Z$-gauge boson given by
Eq.~(\ref{coupling}). As explained in section~\ref{higgsinos}, for
$h_{-} = 0$ there are two Dirac charginos degenerate in mass with two
Majorana neutralinos and the $T$ parameter contribution vanishes. The
mass difference between the neutralinos and charginos grows linearly
with $h_{-}$, as does the coupling of the lightest neutralino to the
$Z$-gauge boson, and the $T$ parameter grows quadratically with
$h_{-}$, as shown in Fig.~\ref{rho}, for values of $h_{+} = 2$ (solid
line) and $h_{+} = 1.5$ (dashed line).

Moderate contributions to the $T$-parameter are not in conflict with
electroweak precision measurements. Even in the absence of any other
physics at the weak scale the corrections to the $T$-parameter coming
from the neutralino and chargino sector may be largely compensated by
the negative contribution induced by the presence of a heavy Higgs,
which contributes to the $S$ and the $T$ parameters in a way
proportional to the logarithm of its mass,
\begin{eqnarray}
\Delta S = \frac{1}{12 \pi}
\log\left(\frac{m_h^2}{m_{h_{ref}}^2}\right)
\nonumber\\
\Delta T = -\frac{3}{16 \pi c_W^2}
\log\left(\frac{m_h^2}{m_{h_{ref}}^2}\right), \label{STHiggs}
\end{eqnarray}
where $m_{h_{ref}}$ is a reference Higgs mass value.

\begin{figure}[htb]
\centering \epsfysize=12cm \leavevmode \epsfbox{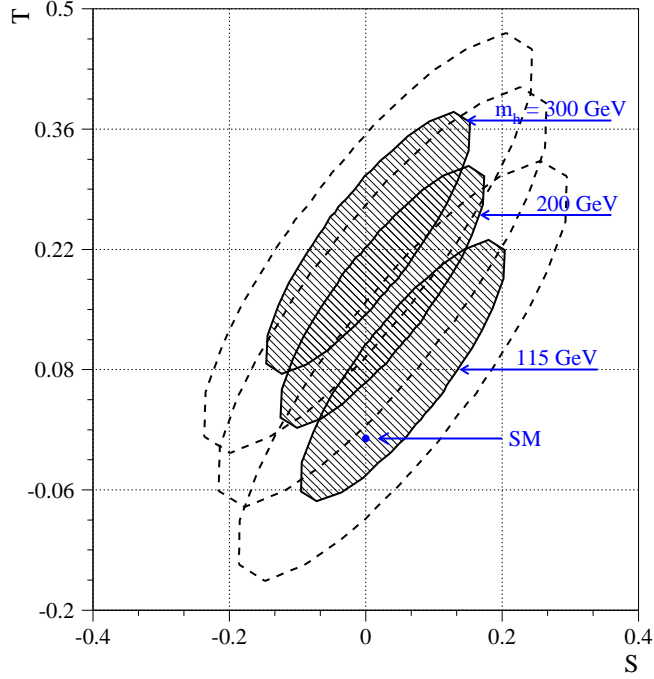}
\caption{ \it 68 \% (solid lines) and \it 95 \% (dashed lines)
C.L. allowed values of the $S$ and $T$ parameter based on the fit to
the precision electroweak data in Ref.~\cite{LEPEWWG}, for three
different values of the Higgs mass: 115~GeV, 200~GeV and 300~GeV.}
\label{stu}
\end{figure}

The model under analysis falls therefore under the class of models
which give a small contribution to the $U$ parameter and sizeable
contributions to the $T$ parameter. Although the new physics gives
only negligible contribution to $S$, a sizeable contribution to the
$S$ parameter may also be induced by a heavy Higgs boson.  A fit to
the precision electroweak data in this class of models has been done
by the LEP electroweak working group~\cite{LEPEWWG}. For a reference
Higgs mass value of 150 GeV, they find
\begin{eqnarray}
S = 0.04 \pm 0.10
\nonumber\\
T = 0.12 \pm 0.10
\end{eqnarray}
with an 85\% correlation between the two parameters. Taking this into
account we show in Fig.~\ref{stu} the 68\% C.L.  (solid lines, hatched
ellipses) and 95\% C.L. (dashed lines) region of allowed values of the
new physics contribution to the $S$ and $T$ parameters for values of
the Higgs boson mass equal to 115~GeV, 200~GeV and 300~GeV,
respectively. Observe that, since we are presenting a fit to the
allowed new physics contribution to the $S$ and $T$ parameters, for
each value of the Higgs mass the origin of coordinates represents the
SM value.
%

From the results of Figs.~\ref{relic},~\ref{rho} and~\ref{stu}, we see
that for a fixed value of the average Yukawa coupling, $h_{+}$, the
requirement of an acceptable relic density and a good fit to the
precision electroweak data implies an interesting correlation between
values of $h_{-}$, the parameter $|\mu|$ and the Higgs mass. The value
of the Higgs mass is also bounded from above from the requirement of a
successful generation of the baryon asymmetry.  For instance, for
$h_{+} = 2$, Fig.~\ref{mhiggs} shows an upper bound on the Higgs mass
of about 200~GeV and therefore from Fig.~\ref{stu} it follows that a
good fit to the electroweak data may only be obtained for values of
the new physics contribution to the parameter $T \simlt 0.27$ (if the
new physics contributions to the parameter $S$ are small, as in the
case under analysis). The upper bound on $T$ translates into an upper
bound on $h_{-}$. From Fig.~\ref{rho} we get that $|h_{-}| \simlt$
0.18, and from Fig.~\ref{relic} this leads to an acceptable relic
density only for $|\mu| \simlt 65$~GeV.

For $h_{+} \simlt 1.6$ the corrections to the precision electroweak
parameters are slightly smaller than for $h_{+} = 2$. However, as
emphasized before, from Fig.~\ref{mhiggs} we see that consistency with
a strong electroweak phase transition can only be obtained for a Higgs
mass value close to its present lower bound, $m_h < 125$~GeV, and
small values of $|\mu| < 100$~GeV. From Fig.~\ref{stu}, this implies
that the new physis contribution to the parameter $T \simlt 0.15$, and
therefore from Fig.~\ref{rho} one obtains that $|h_{-}| \simlt 0.14$.
Interestingly enough, from Fig.~\ref{relic}, we find that the same
region of parameters is consistent with the Dark Matter relic density
provided $|\mu| < 70$~GeV. Observe that the allowed values of
$|\mu|$ approximately coincide with the ones obtained for $h_+ = 2$.

\section{\sc Large Yukawa couplings in low energy supersymmetric theories}
\label{strongc}

In this work we have analyzed the properties of a model with light
charginos and neutralinos which are strongly coupled to the SM
Higgs. Such large values of the Yukawa couplings do not arise in
minimal supersymmetric extensions of the Standard Model, since these
couplings are determined by supersymmetric relations to the weak
couplings. It is therefore an important question to understand under
which conditions such a low energy effective theory may be obtained.

One possibility is to assume that, although the quantum numbers of the
light particles are those of Higgsinos and gauginos of supersymmetric
theories, the theory is not related to any supersymmetric theory at
high energies. A more interesting possibility would be to consider
this theory as a particular realization of split
supersymmetry~\cite{Giudice}, but where the particular relation
between the Yukawa couplings and the gauge couplings has been broken
by supersymmetry breaking interactions.  One of the problems with this
alternative is the one of vacuum stability. In addition, due to the
strong Yukawa couplings, perturbative consistency is lost at scales
much lower than the GUT scale and therefore perturbative unification
of the gauge couplings cannot be achieved.

In this section we shall show that this low energy effective theory
may also arise in low energy supersymmetric models with extra,
strongly coupled gauge sectors, if the scale of supersymmetry breaking
is larger than the scale of spontaneous symmetry breakdown of the
extended gauge sector to the Standard Model one. A supersymmetric
extension provides the necessary stabilizing fields in a natural
way. On the other hand in this extension the strong Yukawa couplings
are proportional to the strong gauge couplings, which become
asymptotically free at high energies. This ultraviolet completion of
the theory allows the preservation of perturbative consistency up to
high energies and therefore the possibility of perturbative
unification of gauge couplings.

In order to illustrate this property let us consider the model
analyzed in Ref.~\cite{Argonne}. The model is based on the low energy
gauge group
\beq
SU(3)_C \otimes SU(2)_1 \otimes SU(2)_2 \otimes U(1)_Y.
\label{gaugeg}
\eeq
First and second generation left-handed quark and leptons transform in
the fundamental representation of $SU(2)_1$, i.e.~$({\bf 2},{\bf 1})$
of $SU(2)_1 \otimes SU(2)_2$, while the third generation left-handed
quark and leptons and the two MSSM Higgs bosons transform in the
fundamental representation of $SU(2)_2$, $({\bf 1},{\bf 2})$. The
model also includes a bifundamental $({\bf \bar 2},{\bf 2})$ Higgs
field $\Sigma$, as well as a singlet field $S$. The scalar component
of the bifundamental takes a vacuum expectation value $\langle
\Sigma\rangle = u \cdot I$, breaking $SU(2)_1 \otimes SU(2)_2 \to
SU(2)_L$. The gauge bosons
\beq
W^{\mu} = \frac{g_2 W_1^{\mu} - g_1 W_2^{\mu}}{\sqrt{g_1^2 + g_2^2}}
\label{WL}
\eeq
remain massless after this symmetry breakdown and interact with an
effective gauge coupling $g_W = g_1 g_2/\sqrt{g_1^2 + g_2^2}$. This
model can be made consistent with gauge coupling unification by
embedding the group $SU(3)_c\otimes SU(2)_1\otimes SU(2)_2\otimes
U(1)_Y$ into the grand unified product group $SU(5)\otimes SU(5)$
broken by bi-fundamental field $({\bf\bar 5},{\bf 5})$ diagonal
VEV's~\cite{Argonne,Kribs}.

In the model of Ref.~\cite{Argonne} there are extra fields
transforming under $SU(2)_1$ but not under $SU(2)_2$. With this
particle content the coupling $g_2$ of $SU(2)_2$ is asymptotically
free, but $g_1$ of $SU(2)_1$ is not. We shall work under the
assumption that the coupling $g_2$ becomes strong at the scale $u$ of
spontaneous symmetry breaking of the symmetry $SU(2)_1 \otimes
SU(2)_2$, and therefore the Winos of $SU(2)_2$ interact strongly with
the Higgs and Higgsinos.  For our analysis here, the relevant term in
the superpotential is $W = M_{\Sigma} \Sigma \Sigma$, and the relevant
supersymmetry breaking terms in the gaugino-Higgsino sector are the
masses of the gauginos, ${\cal{M}}_i$, corresponding to the two groups
$SU(2)_i$. In the absence of supersymmetry breaking terms the
superpartners of $W^{\mu}$ would become the low energy Winos,
interacting with Higgsinos and Higgs bosons with the weak coupling
$g_W$.  In order to get a strongly interacting Wino-Higgsino-Higgs
sector at low energies we need to decouple the bifundamental Higgsinos
$\tilde{\Sigma}$, as well as the weakly interacting Wino $\tilde{W}_1$
from the weak scale theory. This may be achieved by choosing the
supersymmetry breaking parameters ${\cal{M}}_i$ and the Higgsino mass
parameter $M_{\Sigma}$ to verify
\beq
{\cal{M}}_1, \ M_{\Sigma} \gg g_2 u, \qquad
{\cal{M}}_2 \ll g_2 u
\eeq
Supersymmetry is therefore broken before the scale of breakdown of
$SU(2)_1\otimes SU(2)_2$ group to $SU(2)_L$.  The Wino of $SU(2)_1$ is
only weakly coupled with the bidoublet Higgsinos and its large
supersymmetry breaking mass ensures its decoupling from the low energy
theory.

For the parameters given above the low energy Wino has a mass
\beq
M_2 \simeq {\cal{M}}_2 -
\frac{g_2^2 u^2}{M_{\Sigma}}
\label{m2}
\eeq
and has a component on the strongly coupled Wino of $SU(2)_2$ of order
$\cos\theta_{\Sigma}$, with
\beq
\sin\theta_{\Sigma} \simeq g_2 u/M_{\Sigma}.
\eeq
The effective Yukawa couplings between the low energy Winos and the
two Higgsinos are therefore given by
\begin{eqnarray}
h_1 \simeq g_2 \cos\theta_{\Sigma} \cos\beta/\sqrt{2}
\nonumber\\
h_2 \simeq g_2 \cos\theta_{\Sigma} \sin\beta/\sqrt{2}
\end{eqnarray}
where we have assumed that the CP-odd Higgs mass is larger than the
weak scale, and therefore a single, SM-like, CP-even Higgs boson
remains in the low-energy theory.  Assuming $\alpha_2 = g_2^2/4\pi
\simlt 1$ we get that, in this realization of the low energy theory,
the Yukawa couplings $h_i \simlt 2$.

On the other hand, a very large supersymmetric mass $M_{\Sigma}$ would
also demand to be compensated in a precise way by a similarly large
supersymmetry breaking mass in order to get the proper $SU(2)_1
\otimes SU(2)_2$ breaking scale. Lower values of $M_{\Sigma}$ would
reduce the strongly coupled gaugino component of the low-energy Wino
state and would require a fine-tuning between the two terms in
Eq.~(\ref{m2}) in order to obtain a small value of the effective
low-enegy Wino mass. Therefore in this model, a moderate amount of
fine-tuning is required to get consistency with phenomenological
constraints and a strong electroweak phase transition~\footnote{We
thank D.~Morrissey for helpful discussions on this point.}.

We can now ask if the strongly coupled gauge bosons may serve as the
stabilizing bosonic fields defined in section~\ref{phase}.  Let us
first stress that the particular extension of the MSSM presented above
leads to extra contributions to the precision electroweak
observables. Small values of these extra contributions may only be
obtained for values of $g_2 u$ larger than a few
TeV~\cite{Argonne,Liz}. Since $g_2 u$ acts as the bosonic mass $\mu_S$
defined in Eq.~(\ref{bosonstab}), these particles can only act as
stabilizing fields if the Higgs is heavier than the range consistent
with a strongly first order phase transition. It would be interesting
to investigate possible regions of parameter space in which
cancelations between different contributions take place and
consistency with data may be achieved for lighter gauge boson
masses. Otherwise, additional fields would be necessary to stabilize
the Higgs potential. 

A very simple possibility, that has already been pointed out in
section~\ref{precision} is that the stabilizing fields are gauge
singlets, harmless from the point of view of electroweak precision
measurements. These singlets should couple strongly to the Higgs
sector, in order to stabilize the fermionic part of the
zero-temperature effective potential, but should not contribute
strongly to the Higgs quartic coupling since in that case the phase
transition would become weakly first order. A simple example of a
model producing the required effect is given by a singlet
superfield $P$ coupled to the Higgs field doublets as well 
as to a set of $N$ singlet fields $S_i$, with superpotential
\begin{equation}
W=\lambda_1 {\vec{S}}^{\,2} P+\lambda_2 P H_1 H_2+\frac{1}{2}M_S
{\vec{S}}^{\,2} -\frac{1}{2}M_P P^2+\mu H_1 H_2
\label{superp}
\end{equation}
where all couplings and the masses $M_S$ and $M_P$ are positive.  The
absence of a coupling of the singlet fields $S_i$ to the Higgs
doublets as well as the appearence of only terms proportional to
${\vec{S}}^{\,2}$ in the superpotential may be understood as a result
of the invariance of the theory under a global $O(N)$ symmetry.

The supersymmetric masses of the singlet fields, $M_S$ and $M_P$, are
assumed to be much larger than the weak scale, suppressing the mixing
of light singlet fermions with the standard gauginos and Higgsinos.
We shall also assume that there are supersymmetry breaking effects in
the bosonic $\vec S$-sector that prevent the possibility of
integrating out the superfields $\vec S$ from the weak-scale theory
and allow the bosonic $\vec S$-sector to be the stabilizing fields
with mass given by (\ref{bosonstab}). Such supersymmetry breaking
effects should also ensure the preservation of the modifications to
the low-energy Higgs quartic coupling induced by the presence of the
superfields $\vec S$ in the theory.

Furthermore we will assume that the superfield $P$ can be integrated
out supersymmetrically so that for scales below $M_P$ it gives rise to
an effective superpotential as
\begin{equation}
W_{\rm eff}=\frac{1}{M_P}\left(\lambda_1 {\vec{S}}^{\,2}
+\lambda_2 H_1 H_2\right)^2+
\frac{1}{2}M_S {\vec{S}}^{\,2}+\mu H_1 H_2
\label{effsup}
\end{equation}
The above superpotential gives rise to a coupling in the tree-level
potential between the $S$ sector and the Higgs sector, $h_b^2
|\vec{S}|^2|H|^2$, where
\begin{equation}
h_b^2\sim \frac{M_S}{M_P}\lambda_2 \lambda_1 \sin 2\beta ,
\end{equation}
to an $H$-quartic coupling as $\frac{1}{4}\Delta\lambda |H|^4$, where
\beq
\Delta\lambda\sim \frac{\mu}{M_P}\lambda_2^2\sin 2\beta ,
\eeq
and to a self-interacting quartic coupling $\lambda_S |\vec{S}|^4$ fields  
\beq
\lambda_{S} \sim \frac{M_S}{M_P} \lambda_1^2 .
\eeq

Since we have concentrated in this work on the results obtained in the
model with strongly interacting gauginos and Higgsinos for small
values of $h_{-}$, the value of $\tan\beta \simeq 1$, and therefore
$\sin 2 \beta \simeq 1$.  For $M_S \sim M_P$ and $\lambda_2 \lambda_1
\simgt 1$, one obtains values $h_b^2 \simgt 1$, necessary for the
singlets $S_i$ to stabilize the one-loop Higgs potential against the
fermion contribution. On the other hand, for $\mu\ll M_P$, as required
to obtain a good description of the dark matter relic density and
precision electroweak observables, the contribution to the quartic
coupling is not too large. This has implications for the Higgs boson
mass.  In fact, for values of $\tan\beta \simeq 1$, the D-term
contribution to the Higgs mass is small and the Higgs mass may be
raised above (or around) the MSSM values, due to the tree-level
contribution
\beq
m_H^2 \simeq \ 2\Delta\lambda v^2 + {\rm (loop-effects) +
(D-term)}\ .
\eeq
Values of the Higgs mass of order 115--200~GeV may be naturally
obtained for $\lambda_2^2\mu/M_P\simeq$ 0.1--0.3, that appear in this
model for values of $M_P$ of the order of the TeV scale and values of
$\lambda_2$ somewhat larger than one. By choosing the appropriate
value of $N$ and the values of the couplings $\lambda_i$, one can
reach a situation similar to the one described in the previous
sections. A detailed analysis of the parameter space in this
particular model is outside the scope of the present paper.

\section{\sc Conclusions}
\label{conclusions}

In this article we have shown that heavy fermions with strong
couplings to the Higgs fields may induce a strengthening of the
electroweak phase transition and can also provide the proper
CP-violating sources for the generation of baryogenesis.  These heavy
fermions, however, also induce for light Higgs bosons an instability
of the Higgs potential at zero temperature and therefore require an
ultraviolet completion of the theory to recover the consistency of the
low-energy theory. In this work, we have assumed that the heavier,
stabilizing fields have oposite statistics but similar couplings and
number of degrees of freedom as the fermion fields.  The above
properties are then associated with the low energy theory consisting
both of the heavy fermions and the heavier, stabilizing fields.

We have illustrated this possibility by considering a model with TeV
scale Higgsinos and gauginos that may lead to a sufficiently strong
first order electroweak phase transition for values of the Higgs mass
as large as 300~GeV. This is quite different from the results of the
MSSM, in which a light stop is necessary, and the Higgs mass should be
lower than $\sim$ 120~GeV to enhance the strength of the electroweak
phase transition. Also at variance with the case of the MSSM is the
fact that in this scenario the particles that induce a strong first
order phase transition are the same ones responsible for the
generation of the baryon asymmetry at the weak scale.

This model preserves most of the properties of low energy
supersymmetry, including a good Dark Matter candidate. Beyond the
problem of vacuum stability, however, the low-energy Yukawa couplings
deviate from the ones obtained in minimal supersymmetric extensions of
the SM, and their strength spoils the perturbative consistency of the
theory at scales of about a few TeV.  We have shown, however, that the
model may be considered as the low energy effective description of a
gauge extended supersymmetric standard model, in which the strength of
the Yukawa couplings is related to the gauge couplings of an extended
asymptotically free gauge sector, that becomes strongly interacting at
TeV scales.  This ultraviolet completion of the model solves the
strong coupling problem and introduces new heavy particles that, due
to their supersymmetric relations to the Higgsinos and gauginos, tend
to stabilize the Higgs potential in a natural way. The stability of
the potential, however, may only be achieved for values of the heavy
particle masses smaller than about 1 TeV. These relatively small
values of the heavy gauge boson masses lead to large corrections to
the precision electroweak observables, unless cancellations between
different contributions occur.  Alternatively, we also have shown that
the model can be completed with singlets that may stabilize the
zero-temperature effective potential while providing Higgs masses
consistent with present experimental bounds and with the required
strongly first order phase transition.  Finally, we have shown that
this model may be tested by electron electric dipole moment
experiments in the near future.

\section*{\sc Acknowledgments}

\noindent
We would like to thank P.~Batra, A.~Delgado, T.~Tait and D.~Morrissey
for very helpful discussions.  This work was supported in part by the
RTN European Programs HPRN-CT-2000-00148 and HPRN-CT-2000-00152, and
by CICYT, Spain, under contracts FPA 2001-1806 and FPA 2002-00748.
Work at ANL is supported in part by the US DOE, Div.\ of HEP, Contract
W-31-109-ENG-38.  Fermilab is operated by Universities Research
Association Inc. under contract no. DE-AC02-76CH02000 with the DOE.

\end{document}